\documentclass[aps,prd,12pt,preprint,superscriptaddress,nofootinbib,floatfix]{revtex4-1}
\usepackage{latexsym}
\usepackage{amsfonts}
\usepackage[latin1]{inputenc}
\usepackage{subfig}
\usepackage{graphicx}
\usepackage{mathrsfs}
\usepackage{amssymb}
\usepackage{amsmath}
\usepackage{verbatim}
\usepackage[dvips]{color}
\def\D0{\slash\!\!\!\!\!\!\!\!\!\:D0}
\def\bml{B\textrm{-}L}

\begin{document}

\begin{flushright}
preprint FR-PHENO-2012-015\\
\today
\end{flushright}
\vspace*{1.0truecm}

\begin{center}
{\large\bf Natural $Z'$ model with an inverse seesaw and leptonic dark matter}\\

\vspace*{1.0truecm}
{\large L.~Basso, O.~Fischer and J.J.~van~der~Bij}\\
\vspace*{0.5truecm}
{\it Physikalisches Institut, Albert-Ludwigs-Universit\"at Freiburg\\
D-79104 Freiburg, Germany}\\
\end{center}

\vspace*{1.0truecm}
\begin{center}
\begin{abstract}
\noindent 
We consider a model for a $Z'$-boson coupled only to baryon minus lepton number and hypercharge. Besides the usual
right-handed neutrinos, we add a pair of fermions with a fractional lepton charge, which we therefore call leptinos.
One of the leptinos is taken to be odd under an additional $Z_2$ charge, the other even. This allows
for a natural (inverse) seesaw mechanism for neutrino masses. The odd leptino is a candidate for dark matter,
but has to be resonantly annihilated by the $Z'$-boson or the Higgs-boson responsible for giving mass to the
former. Considering collider and cosmological bounds on the model, we find that the $Z'$-boson and/or the extra Higgs-boson can be seen at 
the LHC. With more pairs of leptinos leptogenesis is possible. 

\end{abstract}
\end{center}
\maketitle

\section{Introduction}
\label{sect:intro}
The standard model (SM) gives an excellent description of the known laws of particle physics.
However there are a few facts that it cannot explain. Of importance for this paper are the 
neutrino mass spectrum, in particular the smallness of neutrino masses, the baryon-antibaryon asymmetry in the universe and the presence of dark matter (DM).

Though it appears to be very difficult to explain the mass spectrum of the fermions, there is a mechanism that can 
in principle  make it plausible why the neutrino masses are much smaller than the other masses. The mechanism uses the presence of 
right-handed neutrinos with a large Majorana mass. The mass matrix also contains a Dirac mass. If the Majorana mass is 
much larger than the Dirac mass, one finds after diagonalization one (very) light and one heavy neutrino.   
The method is called  the seesaw mechanism. Several realizations exist, the simplest being the type-I seesaw~\cite {Minkowski:1977sc,VanNieuwenhuizen:1979hm,Yanagida:1979as,S.L.Glashow,Mohapatra:1979ia}.
In this paper we  consider  a somewhat more involved form, which is conventionally called the 
inverse seesaw mechanism in the literature~\cite{Mohapatra:1986bd,Mohapatra:1986aw}.

The presence of heavy neutrinos does not only affect the mass spectrum, it can also provide  a mechanism 
to explain the baryon asymmetry of the universe.
The method is called  ``leptogenesis''~\cite{Fukugita:1986hr} (also see~\cite{Plumacher:1996kc,Buchmuller:2004nz,Davidson:2008bu,Canetti:2010aw} for recent reviews). 
An asymmetry in the lepton sector is produced via $CP$-violating heavy lepton decays. 
Subsequently the asymmetry in the lepton sector is transferred to the baryons by means of electroweak (EW) sphaleron processes~\cite{Kuzmin:1985mm,Khlebnikov:1988sr,Burnier:2005hp}.
Most early papers used neutrino masses at the grand unified scale, however a successful TeV scale leptogenesis is possible 
as well if two such neutrinos are degenerate in mass, thereby enhancing the $CP$-asymmetry parameter. This goes under the name of ``resonant'' leptogenesis~\cite{Flanz:1996fb,Covi:1996wh,Pilaftsis:1997jf,Barbieri:1999ma,Pilaftsis:2003gt,Pilaftsis:2005rv}. In this case, flavour effects have to be taken into account~\cite{Abada:2006fw,Blanchet:2006be,Abada:2006ea,Nardi:2006fx,AristizabalSierra:2009mq}.

Cosmological observations imply the existence of non-baryonic dark matter that drives structure formation on large scales and dominates galactic and extra-galactic dynamics. 
Within the SM there is no candidate for this matter. One  therefore has to enlarge the SM.
The easiest way to explain the dark matter is to postulate the existence
of thermally produced Weakly Interacting Massive Particles (WIMPs), with masses roughly at the weak scale. An additional
unbroken symmetry is postulated, that prevents the WIMPs from decaying.



The latest experimental searches (see, e.g.,~\cite{melbourne}) confirm the SM again at higher energy
scales than before. The data leave little space for modifications. In particular complicated extensions
of the SM lead to phenomenological problems, for instance with flavour changing neutral currents, and tend
to need many fine tunings of parameters.
 Minimal extensions are therefore preferable.
The simplest form to enlarge the gauge group of the SM is to add a single $U(1)$ factor,
which has to be a linear combination of hypercharge and \bml , baryon minus lepton number, if one does not want to enlarge the fermion spectrum.
These are  the so-called ``minimal $Z'$ models'' (see, for example~\cite{delAguila:1995rb,Carena:2004xs,Chankowski:2006jk,Ferroglia:2006mj,Salvioni:2009mt,Basso:2011na}).
The spontaneous symmetry breaking of the extra $U(1)$ factor requires at least a new complex singlet scalar field.
If the coupling of the new gauge group contains a term proportional to \bml , the absence of chiral anomalies 
demands the presence of additional SM-singlet fermions. The presence of right-handed neutrinos, one  per generation,
removes all anomalies~\cite{Jenkins:1987ue,Buchmuller:1991ce,Khalil:2006yi,Huitu:2008gf,Basso:2008iv,Perez:2009mu,Iso:2009ss,Basso:2010pe,Basso:2010yz}.

In these minimal models a type-I seesaw mechanism can be introduced, dynamically generating neutrino masses. The parameters controlling the neutrino masses are compatible with resonant leptogenesis~\cite{Sahu:2004sb,Abbas:2007ag,Blanchet:2009bu,Iso:2010mv}. A  $Z_2$ symmetry can be introduced to provide a stable DM candidate~\cite{Okada:2010wd,HernandezPinto:2011eu,Kanemura:2011mw,Okada:2012sg}. In these models with type-I seesaw, some fine-tuning is required to get two neutrinos almost degenerate and thereby have a successful TeV scale leptogenesis.  The $Z_2$ symmetry is needed for stabilising the DM, but has no further
relation with the neutrino masses.

We will show that the inverse seesaw provides a more natural mechanism to 
explain to neutrino masses, dark matter and baryogenesis at the same time.
Within this framework, 2 new neutrinos per generations are included. Once the mass matrix (of the left-handed and the 2 right-handed neutrinos) is diagonalized, besides the usual light SM-like neutrinos, 3 pairs of naturally quasi-degenerate heavy neutrinos appear, thereby easily implementing the requirements for resonant leptogenesis~\cite{Garayoa:2006xs,Blanchet:2009kk,Blanchet:2010kw}. 

In the context of $U(1)_{\bml}$ extensions of the SM, a model with the inverse seesaw realization exists.
The model contains two extra dileptons, one of which enters  the 
neutrino mass matrix~\cite{Khalil:2010iu}. A $Z_2$ symmetry is advocated that leads to zeroes in the mass matrix.


A disadvantage of this model is that this $Z_2$ symmetry has to be broken, in order to avoid exactly 
zero-mass neutrinos. The breaking mechanism is not present in the tree-level Lagrangian, coming
from non-renormalizable operators. We consider this use of ad-hoc non-renormalizable operators as unnatural. In particular, the arbitrary restriction on the operators to the one desirable for the neutrino masses appears unmotivated.

In this paper we discuss a consistent extension of the SM with a $U(1)$ gauge group, related to \bml\, number, providing
for an inverse seesaw mechanism. 
In contrast to Ref.~\cite{Khalil:2010iu}, we only use renormalizable operators.
The mechanism is natural in the sense that we allow for all renormalizable terms in the Lagrangian,
consistent with the symmetries of the fields. We add pairs of fermions with fractional lepton
number, so-called ``leptinos'' to the Lagrangian. One of them is odd under an additional $Z_2$ symmetry, the other even
as are all ordinary SM particles. Both fermions are needed in order to cancel anomalies.
The  $Z_2$ symmetry, together with the \bml\, charge assignments, restricts the form of the neutrino mass matrix.
At  the same time it stabilises the odd leptino, that becomes the dark matter candidate in the model.

We will present in detail a  version with only one extra pair of leptinos, which is sufficient for the discussion of dark matter.
The possibility of successful leptogenesis requires the extension of the inverse seesaw mechanism to more generations, in order to provide the necessary large phases driving $CP$-violation. The detailed study is beyond the scope of this paper.
 However, we make some comments in the last section and we show that the results concerning the dark matter are not influenced by this extension.


The paper is structured as follows. In the next section, the model is presented. Section~\ref{sect:DM} collects results for
the dark matter abundance generated by the model.
 In section~\ref{sect:leptogenesis} the possibility of a successful leptogenesis is outlined. Finally,  in section~\ref{sect:conclusions}
we present our conclusions. The detailed description of the renormalization group equations (RGEs) of the model is presented for completeness in the appendix.

\section{The model}
\label{sect:model}

We base our extension of the SM on the minimal $Z'$ model~\cite{Carena:2004xs,Chankowski:2006jk,Ferroglia:2006mj,Salvioni:2009mt,Basso:2011na}. The SM gauge group is extended by including a $U(1)$ factor, related to the \bml\, number, with generic mixing with the $U(1)_Y$. A SM singlet complex scalar $\chi$ is required for the spontaneous symmetry breaking of the further $U(1)$ group, thereby providing the $Z'$ boson a mass. The requirement of anomaly cancellation is fulfilled by introducing one right-handed (RH) neutrino per generation. Furthermore, the inverse seesaw mechanism needs extra SM singlet fermions, 
coming in pairs in order not to spoil the anomaly cancellation. Minimally, just one extra pair of fermions is sufficient to provide a DM candidate. 

The classical gauge invariant Lagrangian, obeying the
$SU(3)_C\times SU(2)_L\times U(1)_Y\times U(1)_{B-L}$ gauge symmetry,
can be decomposed as:
\begin{equation}\label{L}
\mathscr{L}= \mathscr{L}_{YM} + \mathscr{L}_s + \mathscr{L}_f + \mathscr{L}_Y \, .
\end{equation}

\subsection{Gauge sector}
In the gauge field basis in which the kinetic terms in $\mathscr{L}_{YM}$ are diagonal~\cite{Basso:2011na}, the covariant derivative reads:
\begin{equation}\label{cov_der}
D_{\mu}\equiv \partial _{\mu} + ig_S
T^{\alpha}G_{\mu}^{\phantom{o}\alpha}  + ig_WT^aW_{\mu}^{\phantom{o}a}
+ig_1YB_{\mu} +i(g_2Y + g_{BL}Y_{\bml})B'_{\mu}\, . 
\end{equation}
This generic model describes a
continuous set of minimal $U(1)$ extensions of the SM, that can be
labelled by the charge assignments of the particles~\cite{Basso:2011na}.
As any other parameter in the Lagrangian, $g_2$ and $g_{BL}$
are running parameters, therefore their values have to be set at some
scale. Special sets  of popular $Z'$ models (see, e.g.~\cite{Carena:2004xs,Appelquist:2002mw}) can be recovered by
imposing relations between $g_2$ and $g_{BL}$ at the EW scale. However such relations can be changed through the renormalization
group running of the coupling constants~\cite{delAguila:1988jz,delAguila:1995rb,Ferroglia:2006mj}. 
The details of the renormalization group equations (RGEs) are presented in the appendix, as they are not central in the argument of the present paper.

For the following study, it is important to focus on measurable observables, one of which being the $Z'$ total width. In the approximation of massless fermions and neutrinos, with $N_\ell$ generations of leptinos, of which just the $CP$-odd leptinos ($S_2$, the DM candidate) are massive, with in first approximation degenerate masses $M_{S_2}$, the total width reads
\begin{equation}\label{eq:Zpwidth}
\Gamma _{Z'} = \frac{M_{Z'}}{12\pi}\left(\left(8 - \frac{N_\ell}{2} + \frac{1}{18} \sqrt{1 - \left(\frac{2M_{S_2}}{M_{Z'}}\right)^2}\right)g_{\bml}^2 + \left(5 - \frac{N_\ell}{8}\right) g_2^2+\left(\frac{13}{2} + \frac{3 - N_\ell}{2}\right)g_2g_{BL}\right)\, .
\end{equation}


Since the $S_2$ particle is only right-handed, its hypercharge is zero (see table~\ref{tab:quantum_number_assignation}), so that its coupling to the $Z'$ boson does not depend on $g_2$. Moreover, in the evaluation of the DM relic abundance with $S_2$ 
as the DM candidate, we will see that a resonant annihilation with the $Z'$ boson is required. In these conditions, $2M_{S_2}\sim M_{Z'}$ and BR($Z'\to S_2 S_2$) $\to 0$. Being around the resonance, the only parameter that can influence the $S_2S_2\to Z'$ partial amplitude, and thus the relic density, is therefore the $Z'$ boson width. The smaller the total width, the higher the relative partial amplitude. Hence, the minimum relic density, when one keeps $M_{Z'}$ fixed, can be obtained by minimizing the $Z'$ width of eq.~(\ref{eq:Zpwidth}) with respect to $g_2$. By direct computation, we obtain that the total $Z'$ decay width is minimized for

\begin{equation}\label{gtmin}
g_2^{min}(g_{BL})=-2\frac{ 16-N_\ell}{40-N_\ell}g_{BL}\, .
\end{equation}

When $N_\ell=1$ (the case discussed here), $g_2^{min}=-10/13\, g_{BL}$. Other important cases are for $N_\ell=3$ (discussed in section~\ref{sect:leptogenesis} in connection with leptogenesis), for which we obtain $g_2^{min}=-26/37\, g_{BL}$, and $N_\ell=0$, where the minimum width is for the SO(10)-inspired $U(1)_\chi$ model, $g_2^{min}=-4/5\, g_{BL}$.

So far, we have neglected that the $Z'$ boson is in general mixed with the SM $Z$ boson. Typical bounds  from LEP-I measurements
 at the $Z$-boson peak require the mixing angle to be less than $\mathcal{O}(10^{-3})$~\cite{Erler:2009jh}.


Further, from a combination of LEP-I and LEP-II data the ratio mass-over-coupling is bounded to be bigger than several TeV.
 In Ref.~\cite{Salvioni:2009mt} the authors reanalysed the LEP data for a model with the same gauge sector as ours,
 while a specific bound for $g_2=0$ can also be found in Refs.~\cite{Carena:2004xs,Cacciapaglia:2006pk}. 

\begin{figure}[h]
  \begin{center}
  \includegraphics[width=0.7\textwidth,angle=0]{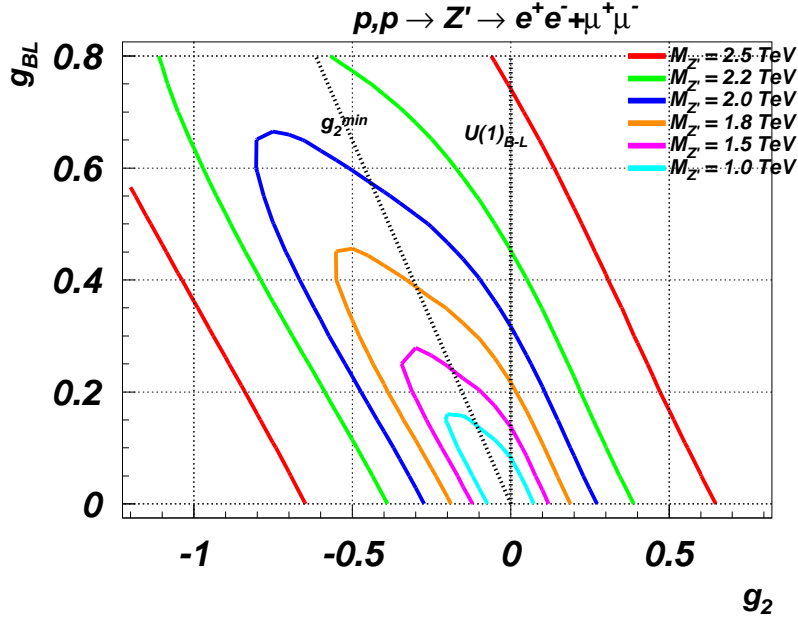}
  \end{center}
  \caption{$Z'$ exclusions from CMS  data, at $\sqrt{s}=7$ TeV for the combination of $4.7$ fb$^{-1}$ in the electron channel and $4.9$ fb$^{-1}$ in the muon channel. The dotted black lines refer to the two main benchmark models of this analysis.}
\label{Zp-excl}
\end{figure}

Before moving on to the results, we present here (see figure~\ref{Zp-excl}) the  $95\%$ C.L. exclusions at the LHC in the $g_{BL}-g_2$ plane, based on the CMS data at $\sqrt{s}=7$ TeV for the combination of $4.7(4.9)$ fb$^{-1}$ in the electron(muon) channels~\cite{CMS-exclusions}. ATLAS has as well published an analysis for $\sim5$ fb$^{-1}$~\cite{ATLAS_exclusions}, but their limits are less tight than the CMS ones. Therefore, we will not present them here.

Table~\ref{tab:exclusions} collects the maximum allowed $g_{BL}$ coupling per given $Z'$ boson mass for the \bml\, model ($g_2=0$) and for the $g_2^{min}$ case of eq.~(\ref{gtmin}), for $N_\ell=1$.

\begin{table}[h!]
	\centering
	\begin{tabular}{|c|c|c|}\hline
$M_{Z'}$ (TeV) & $g_{BL}$ for $g_2^{min}$ & $g_{BL}$ for \bml\, \\
	\hline
2.5 & $>0.8$ & $0.75$ \\
2.2 & $0.81$ & $0.45$ \\
2.0 & $0.58$ & $0.31$ \\
1.8 & $0.39$ & $0.22$ \\
1.5 & $0.24$ & $0.13$ \\
1.0 & $0.13$ & $0.08$ \\
0.5 & $0.05$ & $0.03$ \\\hline
	\end{tabular}
\caption{$95\%$ C.L. exclusions for the benchmark models of interest. Couplings smaller than those in the table are allowed.}\label{tab:exclusions}
\end{table}

\subsection{Scalar sector}
In order to give the $Z'$ boson a mass through spontaneous symmetry breaking, an extra complex scalar $\chi$ is introduced. The scalar Lagrangian reads
\begin{equation}\label{new-scalar_L}
\mathscr{L}_s=\left( D^{\mu} H\right) ^{\dagger} D_{\mu}H + 
\left( D^{\mu} \chi\right) ^{\dagger} D_{\mu}\chi - V(H,\chi ) \, ,
\end{equation}
with the scalar potential given by
\begin{eqnarray} \label{BL-potential}
V(H,\chi ) = m^2H^{\dagger}H + \mu ^2\mid\chi\mid ^2 +
  \lambda _1 (H^{\dagger}H)^2 +\lambda _2 \mid\chi\mid ^4 + \lambda _3
  H^{\dagger}H\mid\chi\mid ^2  \, , 
\end{eqnarray}
where $H$ and $\chi$ are the complex scalar Higgs doublet and singlet
fields, respectively.

The scalar sector is now made of two real {\it CP}-even scalars, a light one $h_1$ and a heavy one $h_2$,
that are mixtures of the Higgs doublet field and the singlet field:
\begin{equation}\label{displ_scalari_autostati_massa}
\left( \begin{array}{c} h_1\\h_2\end{array}\right) = \left( \begin{array}{cc} \cos{\alpha}&-\sin{\alpha}\\ \sin{\alpha}&\cos{\alpha}
	\end{array}\right) \left( \begin{array}{c} h\\h'\end{array}\right) \, ,
\end{equation}
where $v$($v'$) is the VEV of the doublet(singlet) field, while the mixing angle
$-\frac{\pi}{2}\leq \alpha \leq \frac{\pi}{2}$ can be expressed as:\label{displ_scalar_angle}
\begin{eqnarray}\label{sin2a}
\tan{2\alpha} &=& \frac{\lambda _3 vv'}{\lambda _1 v^2 - \lambda _2 (v')^2} \, ,
\end{eqnarray}
with $\lambda_{1,2,3}$ being the parameters entering in the quartic pieces of the scalar potential.
The presence of extra heavy neutrinos and the $Z'$ boson will alter the properties of the Higgs bosons. The scalar mixing angle $\alpha$ is a free parameter of the model, and the light(heavy) Higgs boson couples to the new matter content proportionally to $\sin{\alpha}$($\cos{\alpha}$), i.e. with the complimentary angle with respect to the interactions with the SM particles.

The \bml\,-breaking vev $v'$ can be expressed in terms of the $Z'$ mass and the gauge couplings as follows:
\begin{equation}\label{B-L_vev}
v'=\frac{3M_{Z'}}{2g_{BL}} \sqrt{1-\frac{g_2^2 v^2}{4 M_{Z'}^2-v^2(g^2+g_1^2)}}\, .
\end{equation}
Here we have chosen the \bml\, charge of the complex singlet scalar to be $Y^{\bml}_\chi=\frac{2}{3}$, 
as determined by gauge invariance of the Yukawa sector (see the following section).

The LEP bounds on $M_{Z'}/g_{BL}$ can provide an absolute lower bound on the VEV $v'$ through eq.~(\ref{B-L_vev}). For the benchmark scenarios of interest here, the bounds are:
\begin{eqnarray}\label{LEP_B-L}
g_2=0: &\qquad & 	v' > 9 \mbox{ TeV}\, ,\\ \label{LEP_gtmin}
g_2^{min}: &\qquad & 	v' > 6.7 \mbox{ TeV}\, .
\end{eqnarray}

\subsection{Yukawa sector}
We describe here the inverse seesaw mechanism and its implementation in our model.
In contrast with the existing literature, we employ renormalizable operators {\it only}. The inverse seesaw mechanism can be implemented by means of $N_\ell$ extra pairs of RH fields ($S_1$ and $S_2$) beside the usual $3$ RH neutrinos ($\nu_R$), the latter required by the anomaly cancellation. Therefore,
the Yukawa Lagrangian for the neutrino sector reads
\begin{equation}\label{vdBinverse_lagr}
 \mathcal{L}^\nu_Y= -y^{D}\overline {\ell_{L}} \widetilde H\nu _{R} -y_i^{N}\overline {\nu_{R}^c} S^i_1 \chi  -y_{ij}^{s1} \overline {(S^i_1)^c} S^j_1 \chi^{\dagger}
	         - y_{jj}^{s2} \overline {(S^j_2)^c} S^j_2 \chi+  {\rm 
h.c.} \qquad i,j=1\dots N_\ell \, ,
\end{equation}
where $S^j_2$ are the only odd fields under a $Z_2$ symmetry that is introduced to avoid unwanted $S_1$--$S_2$ mixing. The lightest of the $S^j_2$ particles is our DM candidate. A summation over $i,j$ is implied and an identical number of $S_1$ and $S_2$ fields is required by anomaly cancellation.
As a concrete model for the DM study, we focus on $N_\ell=1$ case.
By convention, we choose to implement the inverse seesaw mechanism in the third generation of leptons only. In the following, the formulas will refer to the latter and we drop the ``$i$'' index, unless otherwise specified. In section~\ref{sect:ext_n_fam} we will show that the DM results do not depend substantially on $N_\ell$.

The gauge invariance of the Yukawa Lagrangian is achieved by solving $Y^{\bml}_{S_1}+Y^{\bml}_{\chi}=-Y^{\bml}_{\nu_R}$ and $2 Y^{\bml}_{S_1}=\pm Y^{\bml}_{\chi}$,
with $Y^{\bml}$ the \bml\, charge of the fields. The following two sets of solutions exist, considering that $Y^{\bml}_{\nu_R}=1$ is fixed by anomaly cancellations:
\begin{itemize}
\item[i)] $Y^{\bml}_{S_1}=-1$ and $Y^{\bml}_{\chi}=2$, that is a replica of the type-I \bml\, model~\cite{Basso:2008iv}. In this case, Majorana masses for all the neutrinos would be allowed, as well as a $\overline {\ell_{L}} \widetilde H S_1$ term. The presence of these terms do not allow for light neutrinos without an extreme fine-tuning of parameters;
\item[ii)] $Y^{\bml}_{S_1}=1/3$ and $Y^{\bml}_{\chi}=2/3$. This is the new solution proposed here, that forbids all the unwanted terms appearing in (i). Because the \bml\, charge is one third of that of the normal leptons and because they do not carry a colour charge, the $S_{1,2}$ fields will be called ``leptinos''  throughout the rest of this paper.
\end{itemize}

Table~\ref{tab:quantum_number_assignation} summarises the particle content and the charge assignments.
 
\begin{table}[!h]
\begin{center}
\begin{tabular}{|c||c|c|c||c|c||c|c|}\hline
$\boldsymbol {\psi} $   & $\ell_L$ & $e_R$ & $\nu _R$ & $S _1$ & $S _2$ & $H$ & $\chi $ \\ \hline 
$\boldsymbol {SU(2)_L}$ & $2$  & $1$   &  $1$ &  $1$  &  $1$  & $2$ & $1$ \\  
$\boldsymbol {Y} $      &  $\displaystyle -\frac{1}{2}$ &  $-1$ &  $0$ &  $0$ &  $0$ & $\displaystyle \frac{1}{2}$ & $0$ \\  
$\boldsymbol {B-L} $  & $-1$ & $-1$ & $-1$ & $\displaystyle\frac{1}{3}$ & $\displaystyle-\frac{1}{3}$ & $0$ & $\displaystyle\frac{2}{3}$ \\  
\hline 
$\boldsymbol Z_2 $  & $+$ & $+$ & $+$ & $+$ & $-$ & $+$ & $+$\\ 
\hline
\end{tabular}
\end{center}\caption{\it $Y$ and \bml\, quantum number and $Z_2$ parity assignations to chiral fermion and scalar fields. \label{tab:quantum_number_assignation}}
\end{table}

After spontaneous symmetry breaking, the two Higgs fields acquire vacuum expectation values (called $v$ and $v'$, respectively). As a consequence, the first 3 terms of eq~(\ref{vdBinverse_lagr}) lead to the following mass matrix for the interacting neutrinos, in the $(\nu^c_l, \nu_R,  S_1)$ basis:
\begin{equation}\label{vdb_inverse_mass_matrix} 
{\mathscr{M}^I} = 
\left( \begin{array}{ccc} 0 & M_D & 0\\ 
                  M^\dagger_D &  0 & M_N\\
		  0  & M_N & M_{S_1}
 \end{array} \right)\, , 
\end{equation} 
where
\begin{equation} 
M_D = \frac{y^{\nu}}{\sqrt{2}} \, v \, , \qquad M_{N} = \frac{y^{N}}{\sqrt{2}} \, v' \, , \qquad M_{S_1} = \sqrt{2}\, y^{s1}\, v' \, .
\end{equation}
Eq.~(\ref{vdb_inverse_mass_matrix}) is generally referred to as the matrix of the inverse seesaw mechanism. Notice that within the model all the terms appearing in the eq.~(\ref{vdb_inverse_mass_matrix}) are proper Yukawa masses. In contrast, in the traditional literature, ${\mathscr{M}^I}_{3,3}$ is an effective mass term parametrizing the lepton number violation required for the neutrinos to become massive. In our model, $L$ is spontaneously broken via \bml .
When $N_\ell=3$ generations of leptinos are considered, $y^\nu$ and $y^{s1}$ are in general $3\times 3$ complex matrices, and it is not restrictive to consider $y^{N}$ as a $3\times 3$ real and diagonal matrix.

After diagonalization of eq.~(\ref{vdb_inverse_mass_matrix}), the neutrino mass eigenstates are called $\nu_l$, $\nu_h$ and $\nu_{h'}$, with $3\times 3$ mass matrices
\begin{eqnarray}\label{neu_masses}
M_{\nu_l} &\sim& M_D M_N^{-1} M_{S_1}(M_N^T)^{-1} (M_D)^\dagger\, , \\ \label{neu_masses2}
M^2_{\nu _h} = M^2_{\nu _{h'}} &\sim& M^2_{D}+M^2_{N}\, .
\end{eqnarray}

However, it is beyond the scope of this paper to consider the most general case in detail. We limit ourselves to the model with only one 
generation of leptinos, which is largely sufficient for the dark matter question. In this case, one can 
without loss of generality choose a basis in which both the 3--component vector $y^{N}$ and the single parameter $y^{s1}$ are real.
The above formulas simplify, since the seesaw mechanism acts only on the third generation, so that $\mathscr{M}^I$ is a $3\times 3$ matrix and eq.~(\ref{neu_masses}) gives the masses of the 3 neutrino eigenstates, where $\nu _h$ and $\nu _{h'}$ combine in a quasi-Dirac fermion.
Regarding the first and second generations, the LH and RH neutrinos obtain the usual Dirac mass term, for which $\mathcal{O}(10^{-12})$ Yukawa parameters are required to fit the light neutrino masses.

We present the analytical solution for the inverse seesaw case for $N_\ell=1$ which is sufficient to describe the main features of the mechanism also when $N_\ell>1$. The mass matrix of eq.~(\ref{vdb_inverse_mass_matrix}) is diagonalized by a $3\times 3$ unitary matrix. Such a matrix can be parametrized by $3$ angles and some phases. The latter will be neglected here.  Formally, the following generic parametrization can be employed:
\begin{equation}\label{BLlin_U}
U=\left( \begin{array}{ccc} 1&0&0\\
                            0&C_{23}&S_{23}\\
                            0&-S_{23}&C_{23}\end{array}\right)
\;
\left( \begin{array}{ccc}   C_{13}&0&S_{13}\\
                            0&1&0\\
                            -S_{13}&0&C_{13}\end{array}\right)
\;
\left( \begin{array}{ccc}   C_{12}&S_{12}&0\\
                            -S_{12}&C_{12}&0\\
                            0&0&1\end{array}\right)\, ,
\end{equation}
where $S_{ij}(C_{ij})=\sin{\alpha_{ij}}(\cos{\alpha_{ij}})$.
In good approximation, the angle defining the mixing between $\nu_R$ and $S_1$ is found to be very close to maximal, i.e., $\alpha_{23} \sim \pi/4$, while the angle parametrizing the mixing between $\nu_L$ and $S_1$ is very small: $S_{13} \sim M_D M_{S_1} / M_{N}^2 \ll 1$.
The unitary matrix of eq.~(\ref{BLlin_U}) can then be simplified to
\begin{equation}\label{BLlin_U_simpl}
U \simeq \left( \begin{array}{ccc} C_{12}&S_{12}&0\\
       -\frac{S_{12}}{\sqrt{2}}&+\frac{C_{12}}{\sqrt{2}}&\frac{1}{\sqrt{2}}\\
       \frac{S_{12}}{\sqrt{2}}&-\frac{C_{12}}{\sqrt{2}}&\frac{1}{\sqrt{2}}
  \end{array}\right)\, ,
\end{equation}
where $\displaystyle S_{12}\sim \frac{M_D}{M_{N}}=\frac{y_Dv}{y_Nv'}$ controls the mixing between $\nu_L$ and $\nu_R$. Altogether:
\begin{equation}
\left( \begin{array}{c} \nu _l \\ \nu_h \\ \nu^\prime_h \end{array} \right)
= U \left( \begin{array}{c} \nu _L \\ \nu_R \\ S_1\end{array} \right)
\end{equation}

Neglecting intergenerational mixing, eq.~(\ref{neu_masses}) can be rewritten as
\begin{equation}\label{neumass_simpl}
M_{\nu_l} \sim M_D M_N^{-1} M_{S_1}(M_N^T)^{-1} (M^\ast_D)^T
  = \sqrt{2}\, \frac{y_{s1}y^2_D}{y^2_N} \, \frac{v^2}{v'}\, .
\end{equation}
To obtain light neutrino masses compatible with experiments (i.e., $M_{\nu_l} < 1$ eV), considering $v' = 10$ TeV, eq.~(\ref{neumass_simpl}) simplifies to
\begin{equation}
 y_{s1}\left( \frac{y_D}{y_N}\right)^2 < 10^{-10}\, .
\end{equation}
A possible solution, which allows for the heavy neutrinos in the $\mathcal{O}(100)$ GeV range (suitable for the LHC), requires $y_N\sim \mathcal{O}(10^{-2})$. Then, $y_D\sim \mathcal{O}(10^{-5})$ implies $y_{s1} < \mathcal{O}(10^{-3})$.


A fundamental difference with the model of Ref.~\cite{Khalil:2010iu} exists in the non-interacting neutrino sector, the $S_2$ fields (here we consider only one of them).
In eq.~(\ref{vdBinverse_lagr}) it is clear that the $S_2$ field acquires a mass after the $U(1)_{B-L}$ symmetry breaking when $\chi$ gets a vev (called $v'$):
\begin{equation}\label{DM_mass}
M_{S_2}=\sqrt{2}\, y_{s2}\, v'\, .
\end{equation}
Therefore, its mass is expected to be of $\mathcal{O}(1)$ TeV, if also $v'$ is at TeV scale. The $Z_2$ symmetry protects this particle from decaying, making it a suitable dark matter candidate. 

{
We have now all the elements to assess how our model compares to the literature. The most important aspect is that an appropriate neutrino sector is derived combining the extension of the SM gauge group and the inverse seesaw mechanism for the first time in a fully renormalisable way. No new mass scale is required, just the one set by the spontaneous breaking of the new \bml\, symmetry, as opposed to the traditional literature for the inverse seesaw mechanism. When this scale is set at the TeV scale, a rich phenomenology for the LHC appears: new particles (the $Z'$-boson, two scalar bosons and several heavy neutrinos) and new signatures (for the heavy neutrino signals, see, e.g., Refs.~\cite{delAguila:2008cj,Basso:2008iv,Hirsch:2009ra}). Our model also provides a solution for the DM and the baryogenesis problems. In the first case, the same $Z_2$ symmetry that protects the zeros in the neutrino mass matrix of eq.~(\ref{vdb_inverse_mass_matrix}) also stabilises the lightest $S_2$ field, that is our DM candidate. The study of the parameter space in which the latter is a viable candidate (i.e., its relic abundance matches the experimentally observed one) is presented in the following section. The baryogenesis problem is solved via the so-called resonant leptogenesis at the TeV scale when large phases in the Dirac neutrino masses are considered. These produce a large $CP$ asymmetry, explaining therefore the observed baryon-antibaryon asymmetry, see section~\ref{sect:leptogenesis}. Notice that this mechanism is naturally resonant in the inverse seesaw case, where no fine tuning is required.
}



\section{Results: dark matter}
\label{sect:DM}
\begin{figure}[h]
  \begin{center}
  \includegraphics[width=0.7\textwidth,angle=-90]{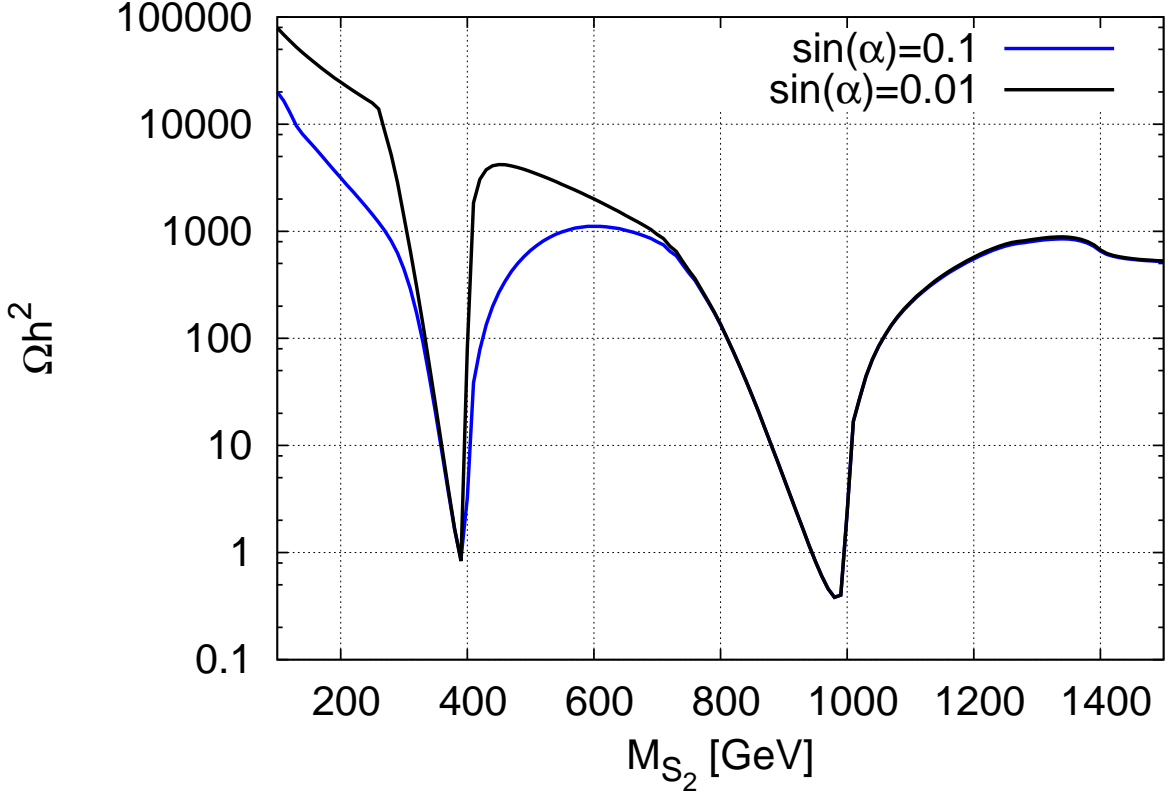}
  \end{center}
  \caption{DM relic abundance as a function of the candidate mass for $M_{H_2}=800$~GeV and $M_{Z'}=2$~TeV, for two different choices of the scalar mixing angle. Also, $g_{BL}=0.1$ and $g_2=0$.}
\label{fig0}
\end{figure}

The $S_2$ field as defined in the previous section is a suitable candidate for WIMP-like cold dark matter  because it is odd with respect to the $Z_2$ symmetry and thus stable on cosmological time scales. An abundance of $S_2$ particles which is thermally produced in the early universe will survive after freeze-out from the thermal bath until today to make up the observed dark matter in the universe.

To compute the relic density of $S_2$ dark matter we used the program MicrOMEGAs \cite{Belanger:2001fz,Belanger:2004yn}, in which the model has been implemented via LanHEP~\cite{Semenov:1996es}. The remaining numerical analysis has been performed in CalcHEP~\cite{Pukhov:2004ca}. The a priori unknown Yukawa couplings of the extra neutrino fields turn out to be negligible in the calculation of the relic density. For concreteness, in the following we fixed  $y^D_{\nu 3}=10^{-4}$, $y^M_{\nu 3}=0.06$ and $y^{S_1}_3=10^{-5}$, owing to $m_{\nu l}=1.5\cdot 10^{-10}$ and $M_{\nu_h}=M_{\nu_h'}=636$ GeV for $v'=15$ TeV.  However, the results shown in the following are insensitive to their precise value.

Our free parameters are $M_{Z'}$, $M_{H_2}$, $\sin{\alpha}$, $g_{BL}$, $g_2$ and the dark matter mass $M_{S_2}$. For a first assessment of the situation, we compute the relic density as a function of the $S_2$ mass for a choice of all other parameters, as shown in figure~\ref{fig0}. 





Comparing the resulting graph with the observed abundance of dark matter:
\begin{equation}
\Omega_{DM} h^2 = 0.1117^{+0.0053}_{-0.0055}\, ,
\label{eq:DM_relic_density}
\end{equation}
%
we conclude that for the $S_2$ particle to be the dark matter, it must annihilate efficiently via a resonant heavy Higgs or via the $Z'$ boson. Also the precise value of $\alpha$ is not relevant in the resonant regions, that are the only important regions for our analysis, as long as it is small, $\sin{\alpha} < 0.1$, as expected if one relies on the confirmation of the recent discovery at the LHC~\cite{LHC_discovery} of a light SM-like Higgs boson. In the following we will study these two different resonant mechanisms fixing $\sin{\alpha}=0.1$ for concreteness, unless otherwise specified because the results do not depend substantially on it.

\subsection{Higgs boson resonant annihilation}

\begin{figure}[!ht]
\centering
\begin{minipage}{0.49\textwidth}
  \includegraphics[width=0.7\textwidth,angle=-90]{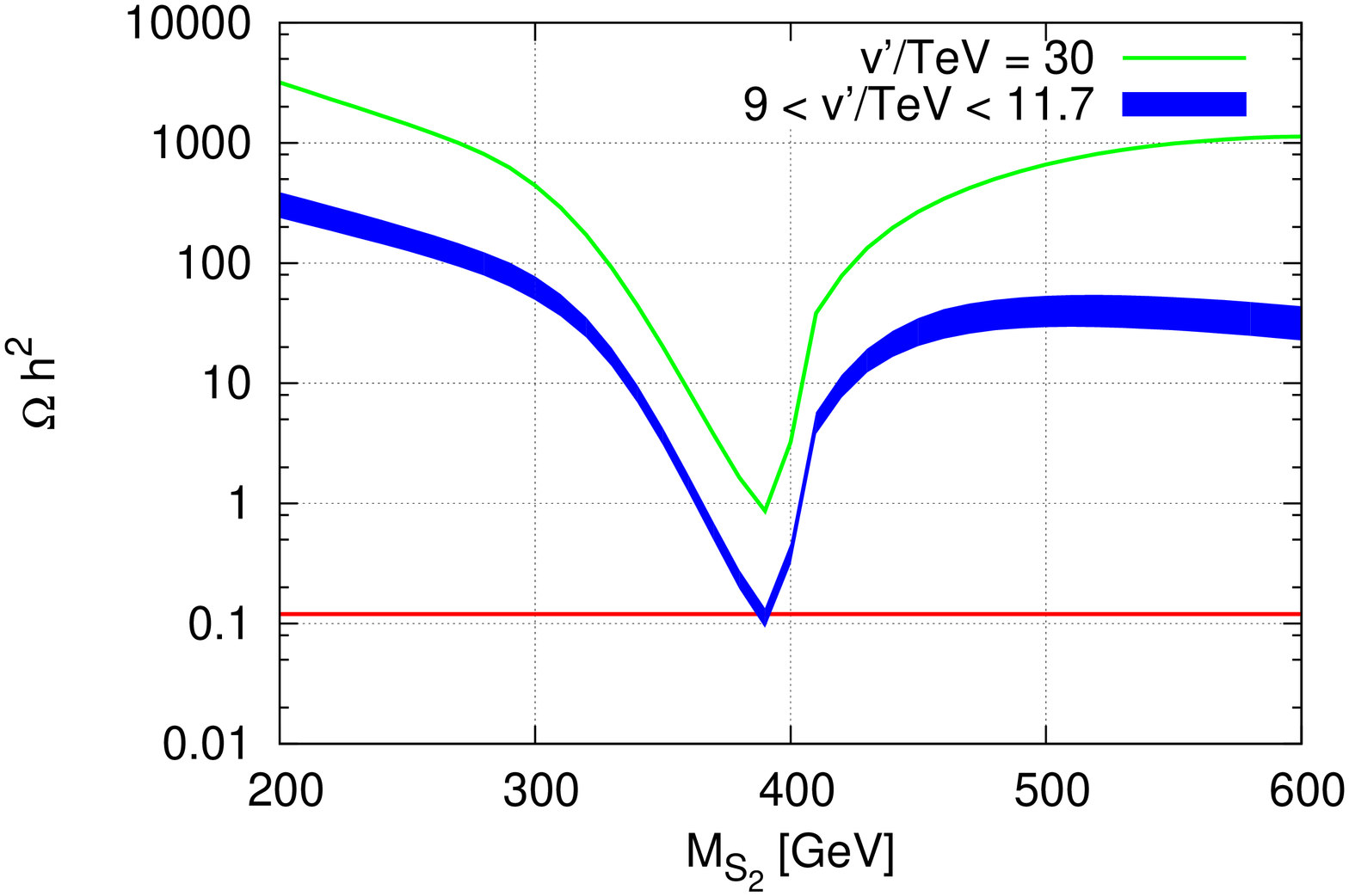}
\end{minipage}  
\begin{minipage}{0.49\textwidth}
  \includegraphics[width=0.7\textwidth,angle=-90]{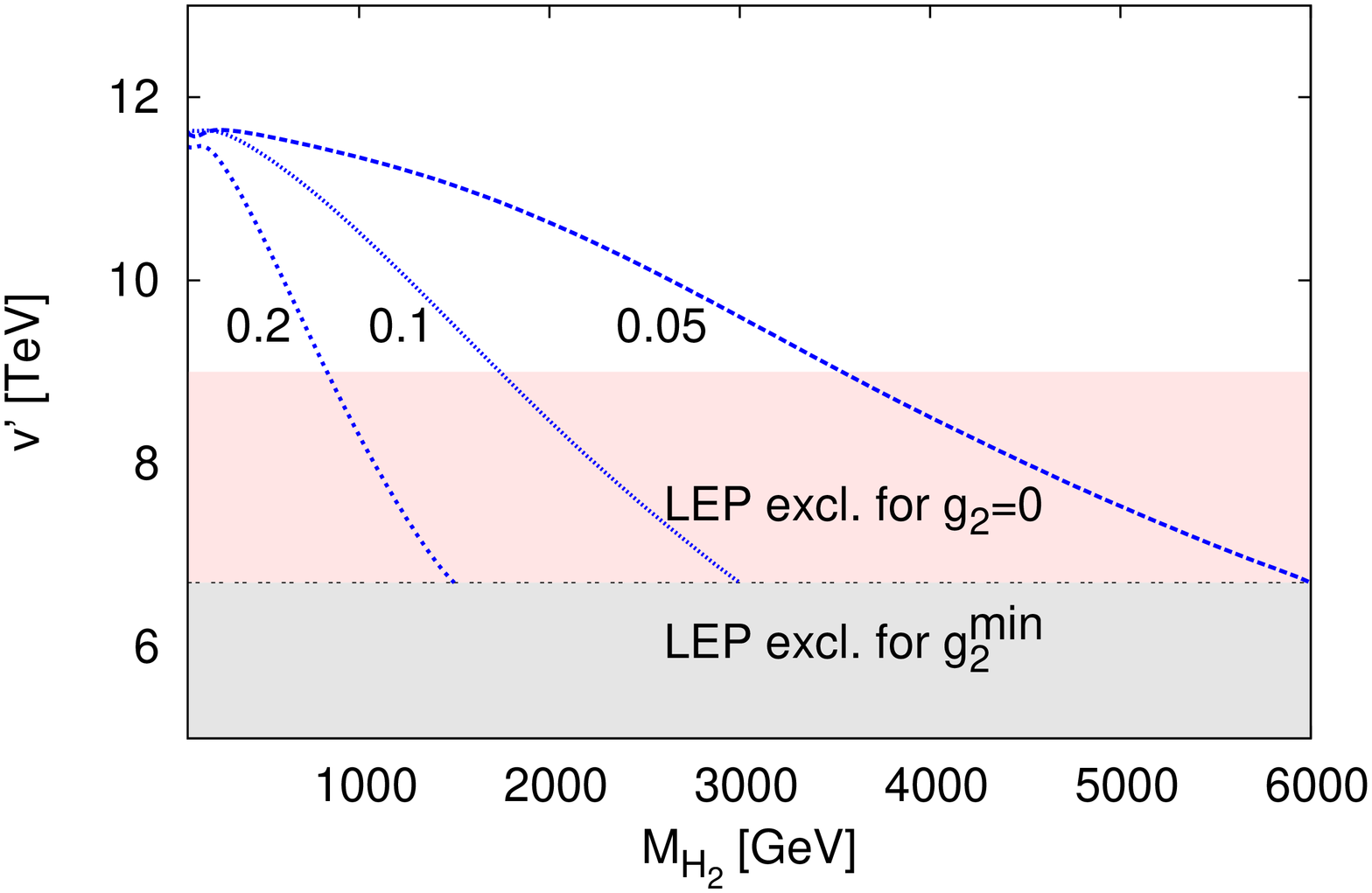}
\end{minipage}
  \caption{({\it Left}) DM relic abundance as a function of the DM mass for $M_{H_2}=800$ GeV, for some values of the vev $v'$ and $\sin{\alpha}=0.1$. The blue shading represents values of the vev for which an allowed DM mass exists, for this $M_{H_2}$, as taken from the right panel.\\
({\it Right}) Allowed region for $v'$ for which a DM mass yields the correct relic density, as a function of the heavy Higgs boson mass, for the three values of $\sin{\alpha}=0.05,\, 0.1,\, 0.3$. The LEP exclusion is as in eqs.~(\ref{LEP_B-L})--(\ref{LEP_gtmin}).}
\label{fig1}
\end{figure}

For small DM masses, where the main annihilation channel is via the heavy Higgs boson, the relic density is proportional to the $\bml$-breaking vev, since $\Omega h^2 \propto \frac{1}{\sigma} \propto \frac{1}{y^2_{s2}} \propto v'^2$, via eq.~(\ref{DM_mass}). According to LEP, the vev is constrained from below as in eqs.~(\ref{LEP_B-L})--(\ref{LEP_gtmin}) for the models of interest. The left panel of figure~\ref{fig1} shows that for $M_{S_2} \simeq M_{H_2}/2$ the minimal abundance just matches the observed value.

In the right panel of figure \ref{fig1} we see that the demand, that the resonant $S_2$ annihilation proceeds via the Higgs channel, strongly constrains from above the vev $v'$ as a function of the heavy Higgs mass $M_{H_2}$, $v'\le 11.7$ TeV. It is interesting to note that this upper bound on $v'$ is independent of the scalar mixing angle $\alpha$. At the same time, the heavy Higgs mass is constrained from above depending on the value of $\alpha$. For instance for $\sin{\alpha}=0.1,\, M_{H_2}\le 1.8(3.0)$ TeV for $g_2=0(g_2^{min})$.


\subsection{$Z'$ boson resonant annihilation}\label{sect:Zp_res}

Due to LHC direct searches, the $Z'$ boson mass has to be above $2\div 2.5$ TeV, depending on the gauge mixing coupling $g_2$. For the $Z'$ annihilation mechanism to be effective, as already stated, a resonant condition has to be matched, meaning that the DM candidate has to be rather heavy. However, the resonance decay can still be very effective.

\begin{figure}[h]

  \begin{center}
  \includegraphics[width=0.6\textwidth,angle=-90]{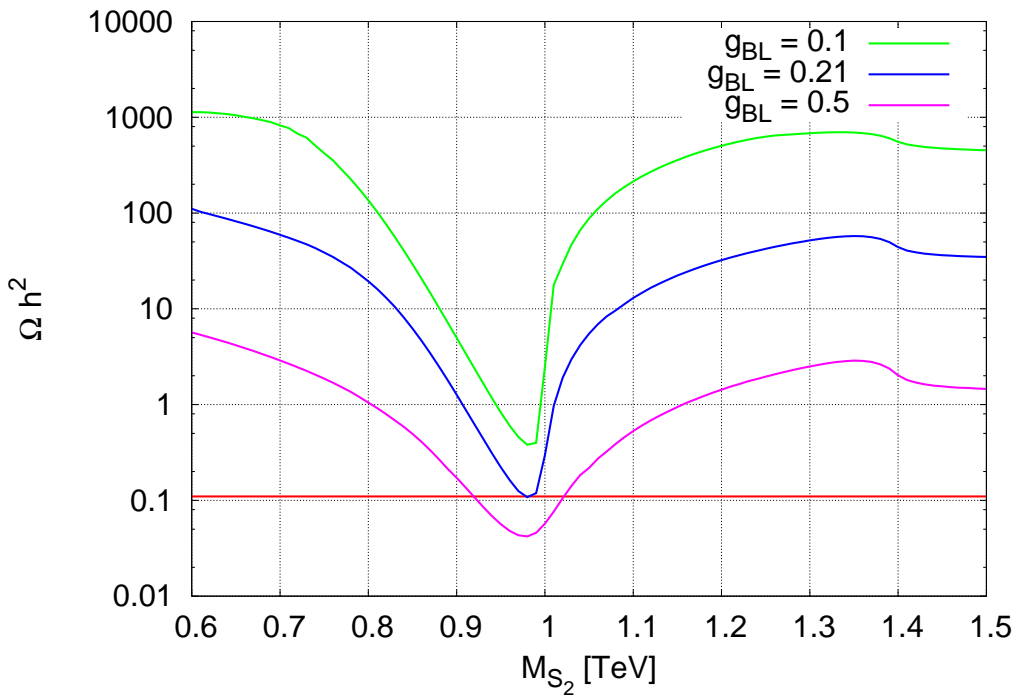}
  \end{center}
  \caption{Relic density as a function of the DM candidate mass around the $Z'$ peak. The lower curve, for $g_{BL}=0.5$, is for illustrative purposes only, being already excluded by LEP.}
\label{fig2}
\end{figure}

Figure~\ref{fig2} shows in more detail that only around the $Z'$ resonance the annihilation is sufficient to match the abundance constraint. For simplicity, we fix here $g_2=0$, so that the relic density is inversely proportional to the square of the gauge coupling, i.e., $\Omega h^2\propto \frac{1}{(g_{BL})^2}$. The lower limit on the vev $v'$ translates into an upper limit on the coupling, for  a fixed $Z'$ mass: in figure~\ref{fig2} the $Z'$ mass is $2$ TeV, which gives the upper limit $g_{BL} < 0.33$, via eq.~(\ref{B-L_vev}). Then, the demand $\Omega h^2 \simeq 0.1$ gives a lower limit for the gauge coupling, $g_{BL}>0.21$.

\begin{figure}[h]
  \begin{minipage}{0.49\textwidth}
  \includegraphics[width=0.9\textwidth,angle=-90]{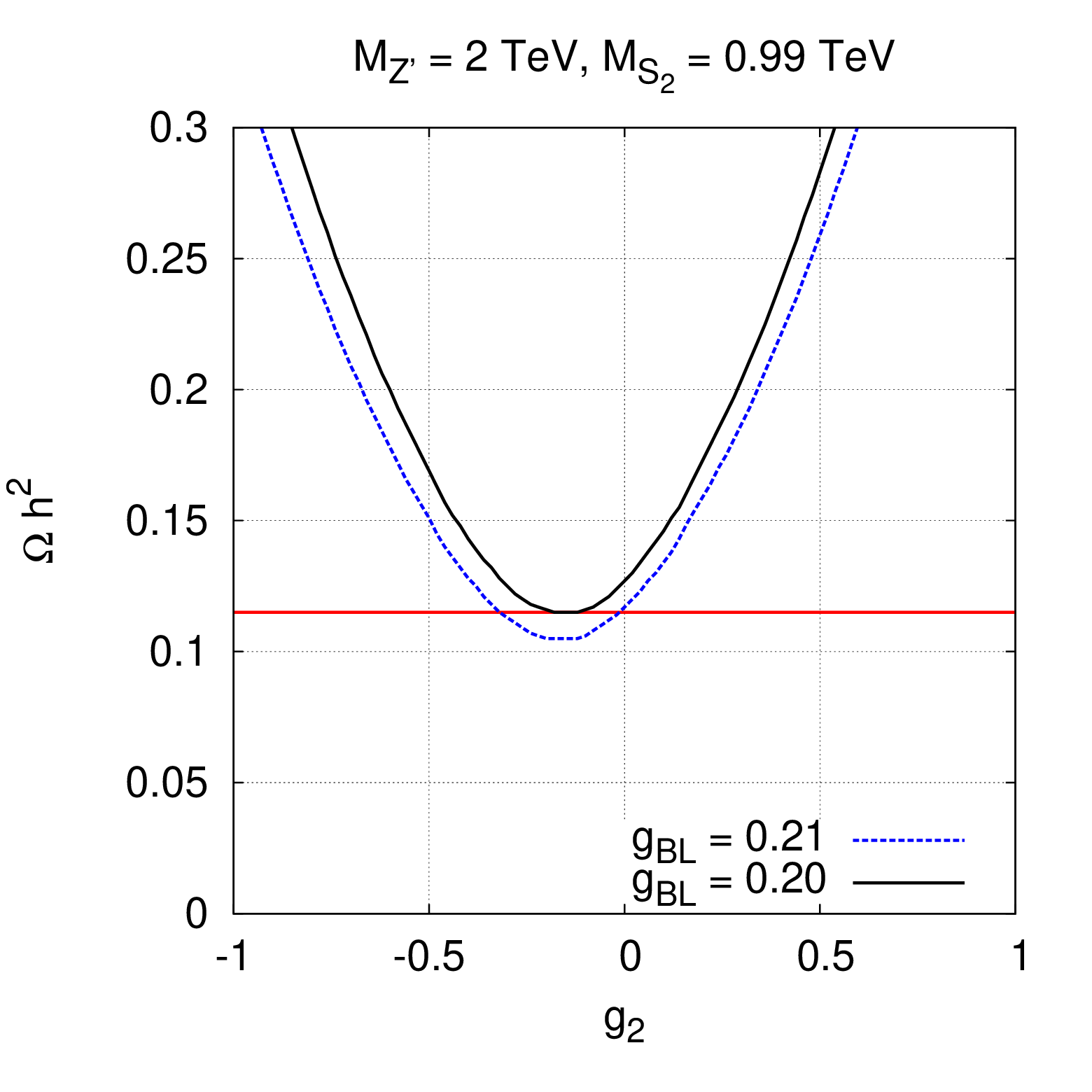}
  \end{minipage}
  \begin{minipage}{0.49\textwidth}
  \includegraphics[width=0.9\textwidth,angle=-90]{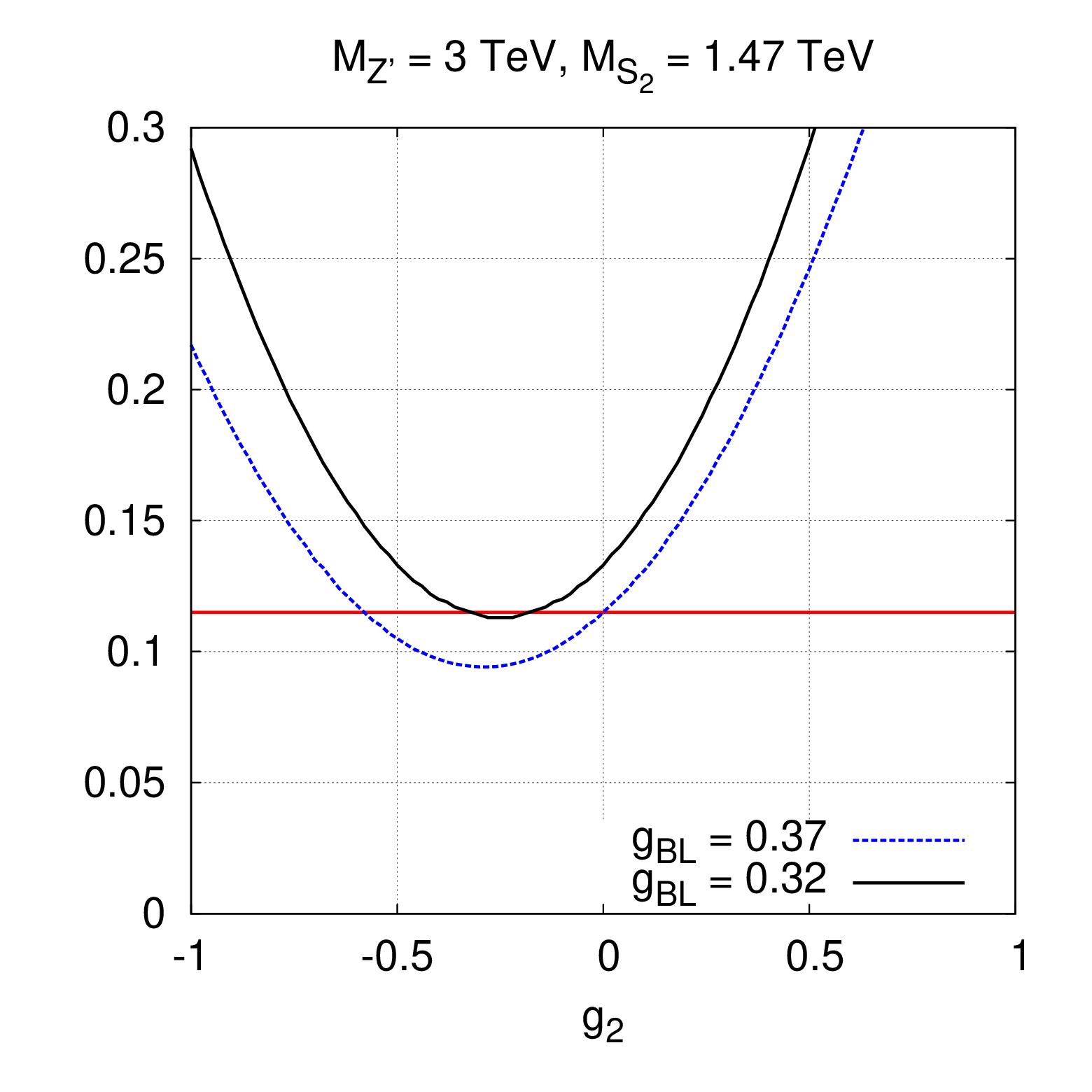}
  \end{minipage}
  \caption{Variation of the relic density (at the $Z'$ resonance) with $g_2$ for a choice of $g_{BL}$ (see the text for further details), for ({\it left}) $M_{Z'}=2$ TeV and ({\it right}) $M_{Z'}=3$ TeV.}
\label{fig3}
\end{figure}

Although the gauge coupling $g_{BL}$ is expected to give the major contribution to the relic abundance evaluation, the impact of $g_2$ might be not negligible either. For fixed $M_{Z'}$  and $g_{BL}$ values, we study here the effect of the mixing gauge coupling. We choose two different sets of $M_{Z'}$ and $g_{BL}$ such that for $g_2=0$ the relic density just satisfies the abundance constraint. Figure~\ref{fig3} then shows that $g_2$ can have an impact on the relic abundance, in particular it can lower the latter when assuming small negative values.

\begin{figure}[h]
\begin{minipage}{0.49\textwidth}
\begin{flushleft}
  \includegraphics[width=0.9\textwidth,angle=-90]{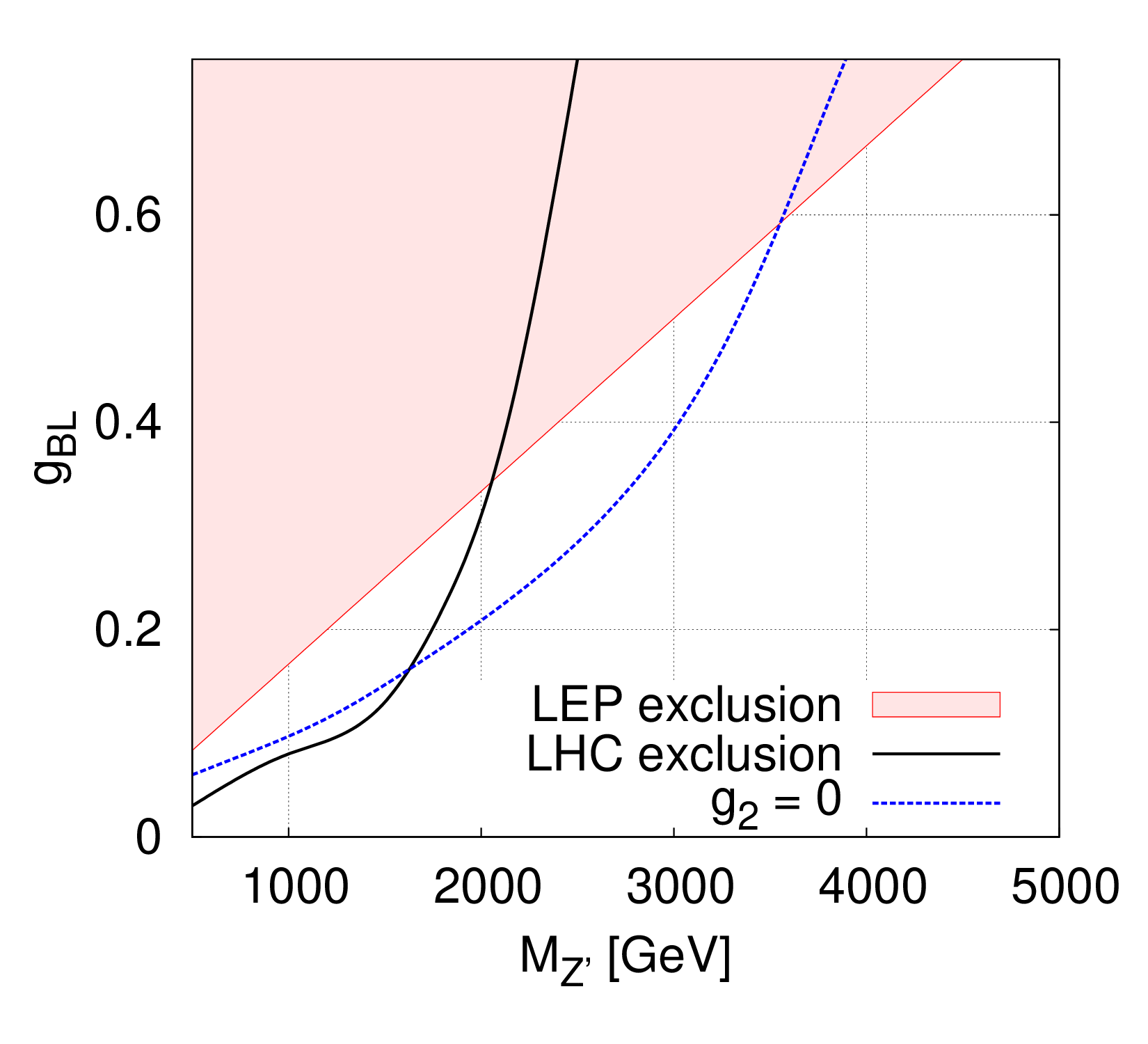}
\end{flushleft}
\end{minipage}
\begin{minipage}{0.49\textwidth}
\begin{flushright}
  \includegraphics[width=0.9\textwidth,angle=-90]{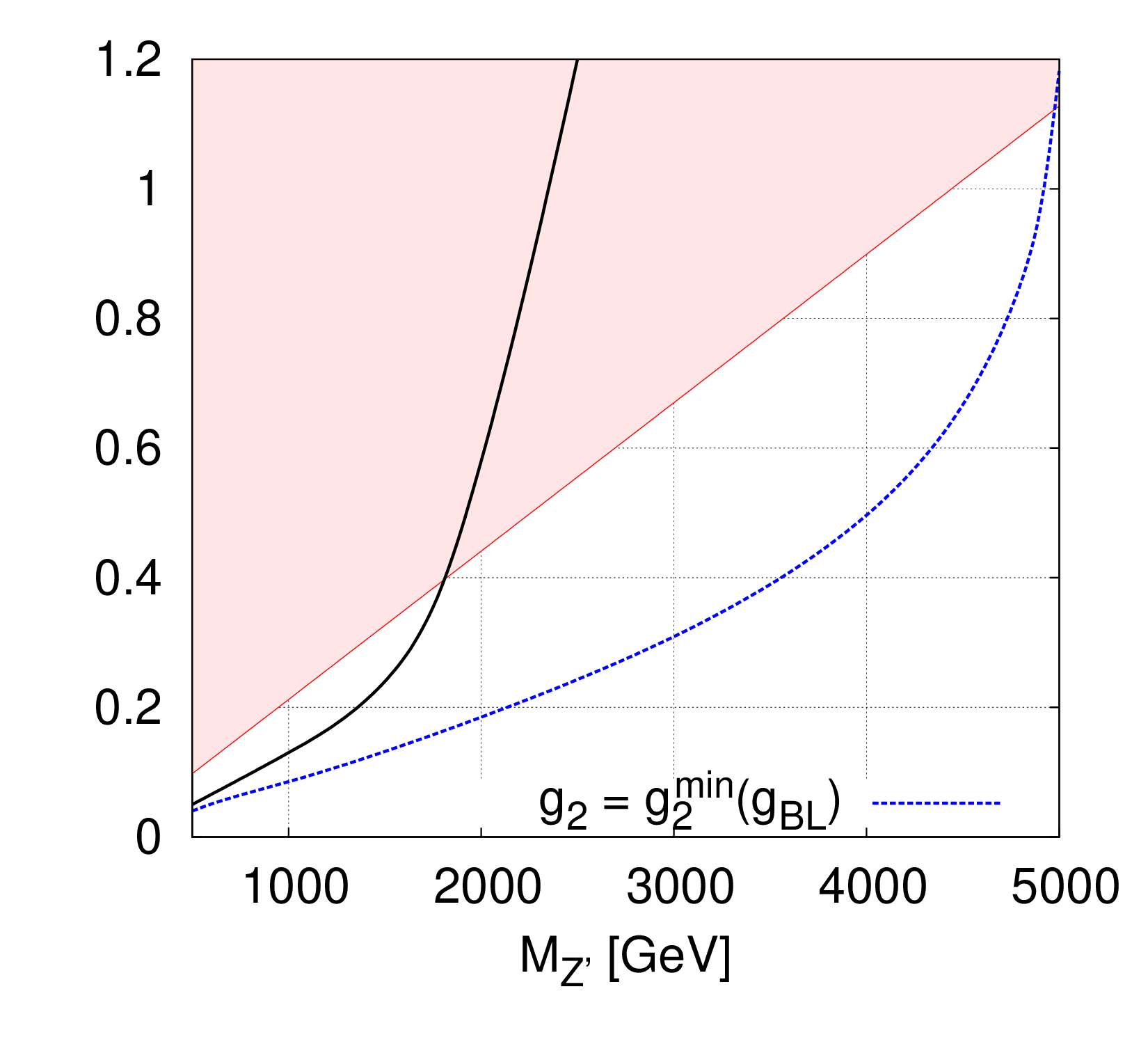}
\end{flushright}
\end{minipage}
\caption{Existence of a suitable DM candidate: the allowed region is the one above the dashed curves, in the $g_{BL}-M_{Z'}$ plane, (left) for $g_2=0$ and (right) for $g_2=g_2^{min}(g_{BL})$, which minimizes the $Z'$ width, hence allowing for a smaller $g_{BL}$ (and therefore a smaller cross section) per fixed $M_{Z'}$. The red shading combinations are forbidden by LEP (eqs.~(\ref{LEP_B-L})--(\ref{LEP_gtmin}), respectively), the black (solid) lines are the LHC exclusion, as in table~\ref{tab:exclusions}.}
\label{figX}
\end{figure}

A general feature is that $\Omega(g_2)$ is growing for $|g_2| \to 1$, so that there exists a value of $g_2$ that minimizes $\Omega(g_2)$.
%
%
%
%
This value of $g_2$ corresponds to the $g_2^{min}$ of eq.~(\ref{gtmin}), that also minimizes the $Z'$ width.
Fixing $g_2$ to $g_2^{min}(g_{BL})$ returns a minimum value for $g_{BL}$ such that the abundance constraint can be matched, which is roughly $5\%$ lower than the lower limit on $g_{BL}$ in the case $g_2=0$.
We can now study the relic density as a function of the $Z'$ boson mass. This is done in figure~\ref{figX}, that shows the range of allowed values for $g_{BL}$ as a function of $M_{Z'}$. The curves, for both $g_2=0$ and $g_2=g_2^{min}(g_{BL})$, limit the existence of a DM candidate mass with suitable relic density above the curves themselves.

As a result, the $Z'$ mass is constrained to $M_{Z'}\leq 3.5(5.0)$ TeV for the 
$g_2=0$ and the $g_2=g_2^{min}(g_{BL})$ case, respectively, which also imposes an upper bound on the gauge couplings when combined with the exclusion limits from LEP.
Notice that this value for the $Z'$ mass is within the LHC ultimate reach~\cite{Basso:2010pe}.
Since we must be on a resonant regime, the upper bound on the $Z'$ mass translates on a upper bound for the $S_2$ mass, $M_{S_2}\leq 2.5$ TeV.

\subsection{Direct detection}

\begin{figure}[h]
\begin{center}
\includegraphics[width=0.5\textwidth,angle=-90]{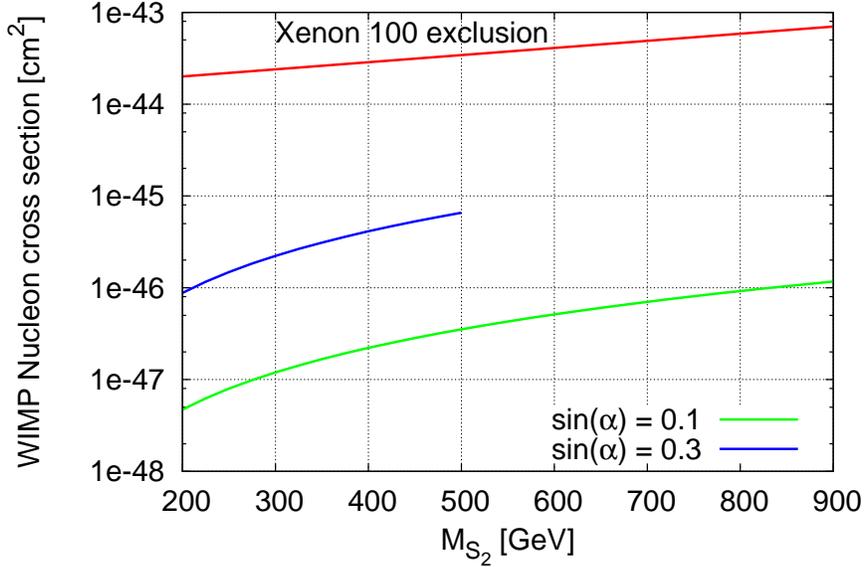}
\caption{Spin-independent direct searches for maximum
Yukawa couplings $y_{S_2}\sim M_{S_2}/v'$. Only allowed masses are plotted, from figure~\ref{fig1}(right), for $g_2=g_2^{min}$.}
\label{fig_direct}
\end{center}
\end{figure}

The spin independent leptino-nucleon interaction is mediated by the light Higgs boson and is therefore sensitive to the value of $(\sin{\alpha}\cos{\alpha})^2$. Figure~\ref{fig_direct} shows that even for the rather large value of $\sin{\alpha}=0.3$, the leptino-nucleon cross section is several orders of magnitude below the actual exclusion limits from XENON100~\cite{Aprile:2011hi}, at most of roughly $10^{-44}$~$\text{cm}^2$ for a DM candidate mass of around $50$~GeV. Since we know that $\sin{\alpha}$ has to be small, also future direct detection experiments will not be able to restrict or to detect the leptino. We  checked that the exchange of $Z'$ bosons consistent with LEP exclusion limits
gives rise to smaller cross sections than Higgs exchange, hence we did not further consider this process.
Even easier to evade are the constraints from spin-dependent leptino-nucleon interaction experiments. Here the only mediator that can play a role is the $Z'$ boson, too heavy for these cross sections to be of any interest.

\subsection{Extension to $N_\ell$ families}\label{sect:ext_n_fam}
\begin{figure}[h]
\begin{minipage}{0.49\textwidth}
  \includegraphics[width=0.9\textwidth,angle=-90]{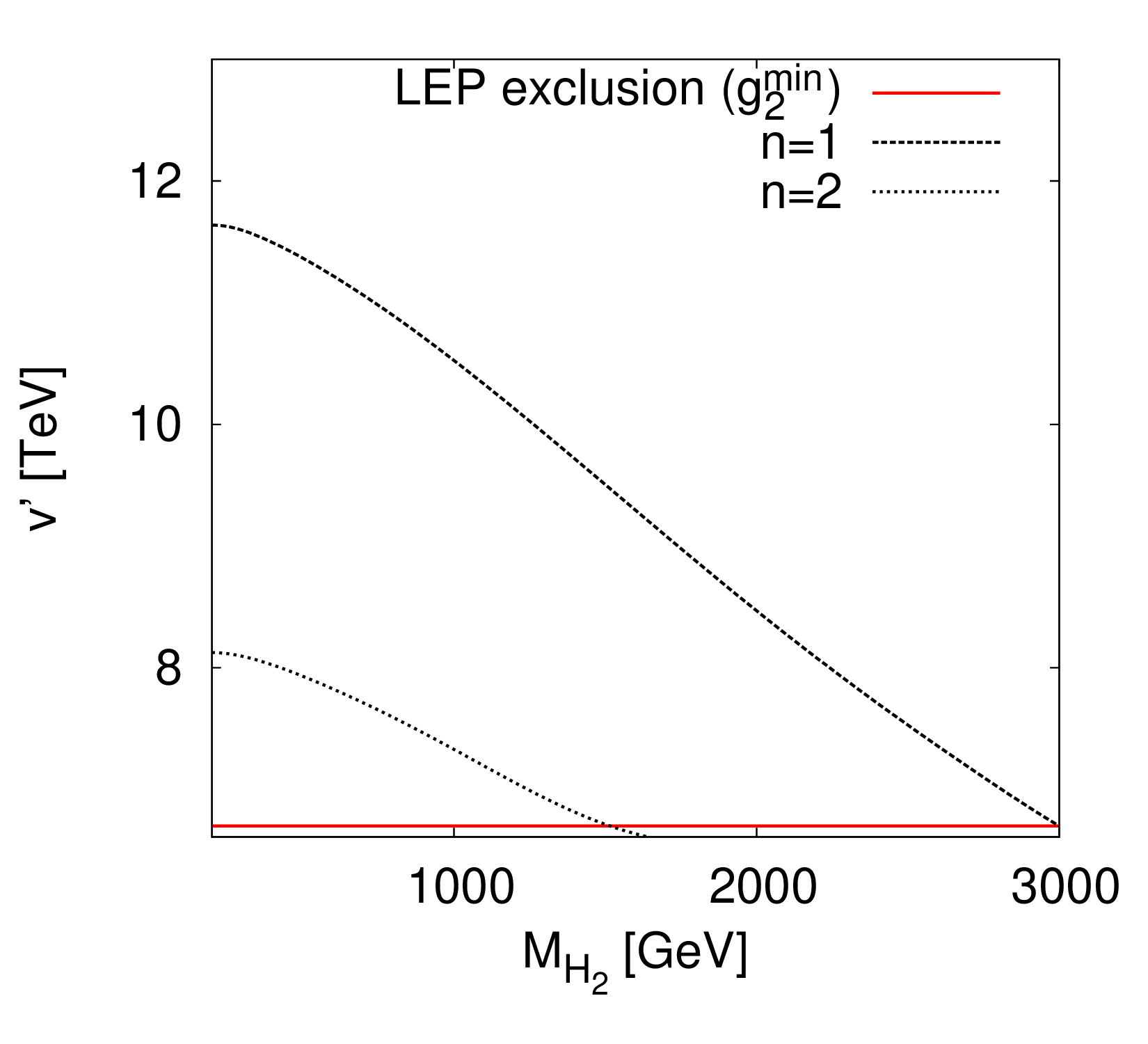}
\end{minipage}
\begin{minipage}{0.49\textwidth}
  \includegraphics[width=0.9\textwidth,angle=-90]{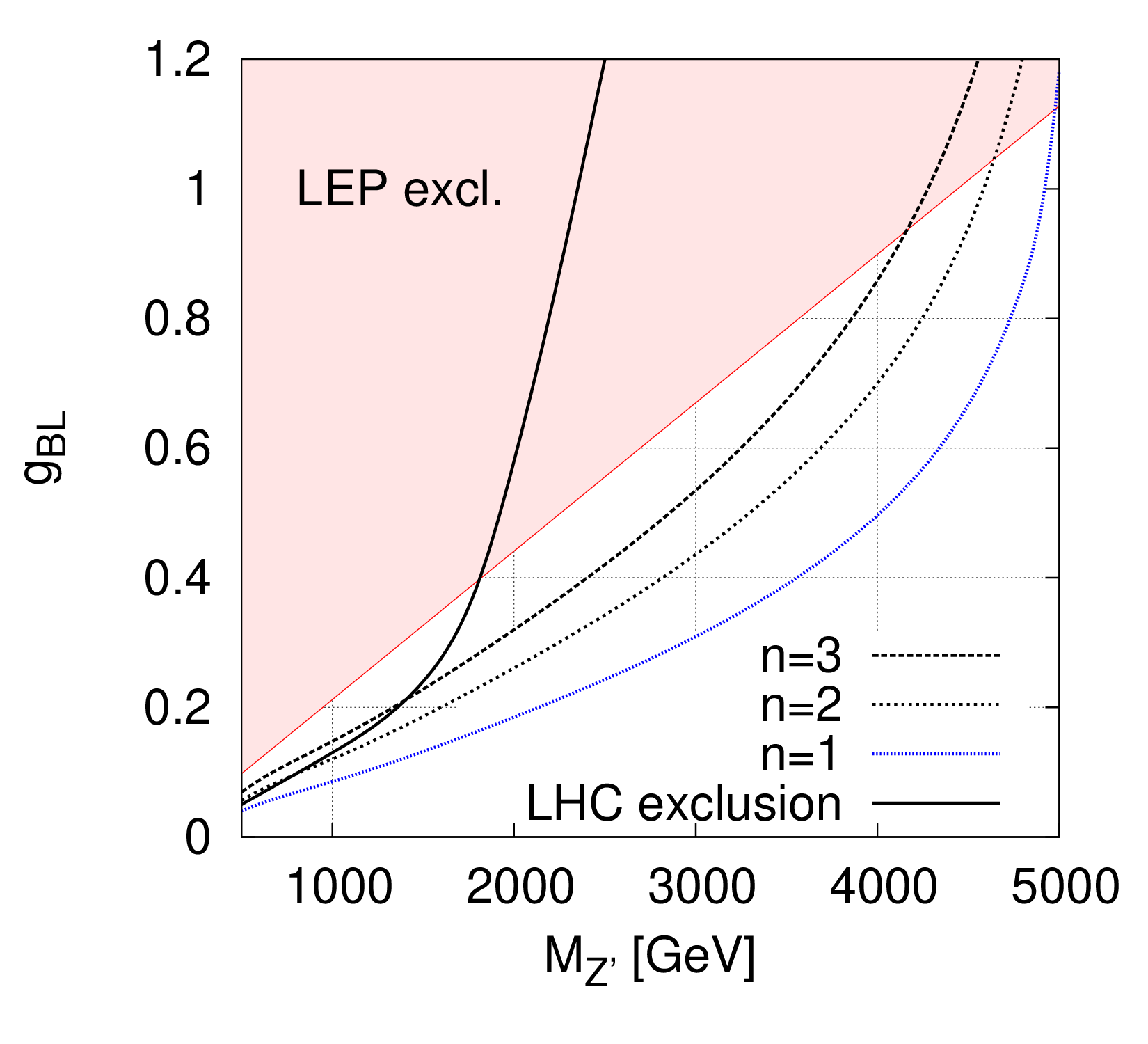}
\end{minipage}
\caption{{\it (Left)} Higgs resonance, $n\equiv N_\ell=3$ is not shown as always disallowed in this case. Here, $\sin{\alpha}=0.1$. {(\it Right)} Allowed parameter range for resonant $Z'$ boson annihilation for $n\equiv N_\ell=1\dots 3$ families of leptinos. The curves are for $g_2=g_2^{min}(g_{BL})$.}
\label{figN}
\end{figure}

As we will discuss in the next section, a successful leptogenesis needs $N_\ell \geq 2$ to provide the required large $CP$--violating phases. Here we want to comment on the impact of extra leptino generations in the DM analysis carried out so far. The extra neutrinos affect the $Z'$ width only, see eq.~(\ref{eq:Zpwidth}), and will be discussed later. First we focus on the leptinos. If the $S_2^i$ fields possess a large mass hierarchy, the heavier particles decouple earlier and decay rapidly to the still thermalized lighter particles. 
If they are exactly mass degenerate, however, we get $\Omega_{N_\ell} = N_\ell \cdot \Omega_1$. Thus to have resonant annihilation the couplings have to be increased by $\sqrt{N_\ell}$. 
Figure~\ref{figN} shows how this affects the parameter space. For resonant heavy Higgs annihilation the vev $v'$ must be reduced by 
a factor $\sqrt{N_\ell}$, so that only the cases $N_\ell=1,2$ can still match the relic abundance for Higgs resonant annihilation, while $N_\ell=3$ is just touching the LEP exclusion limits (for $g_2=g_2^{min}$) from below and therefore such resonant mechanism is not sufficient.

The case of resonant $Z'$ annihilation works well also for $N_\ell=3$. However, the allowed parameter space becomes tighter for higher $N_\ell$. These results are valid when all the leptinos are mass degenerate, and  represents the worst possible case. All others, i.e. when just 2 are degenerate or with a tight mass hierarchy, will be somewhere between the case of $1$ generation and the
case of $3$ generations exactly degenerate in mass. 
Notice that when the leptinos possess a tight mass hierarchy also co-annihilation processes become important and have to be taken into account. However their inclusion will not change the results.

\begin{figure}[!htb]
  \subfloat[]{ 
  \label{Pi_8_2}
  \includegraphics[angle=0,width=0.49\textwidth ]{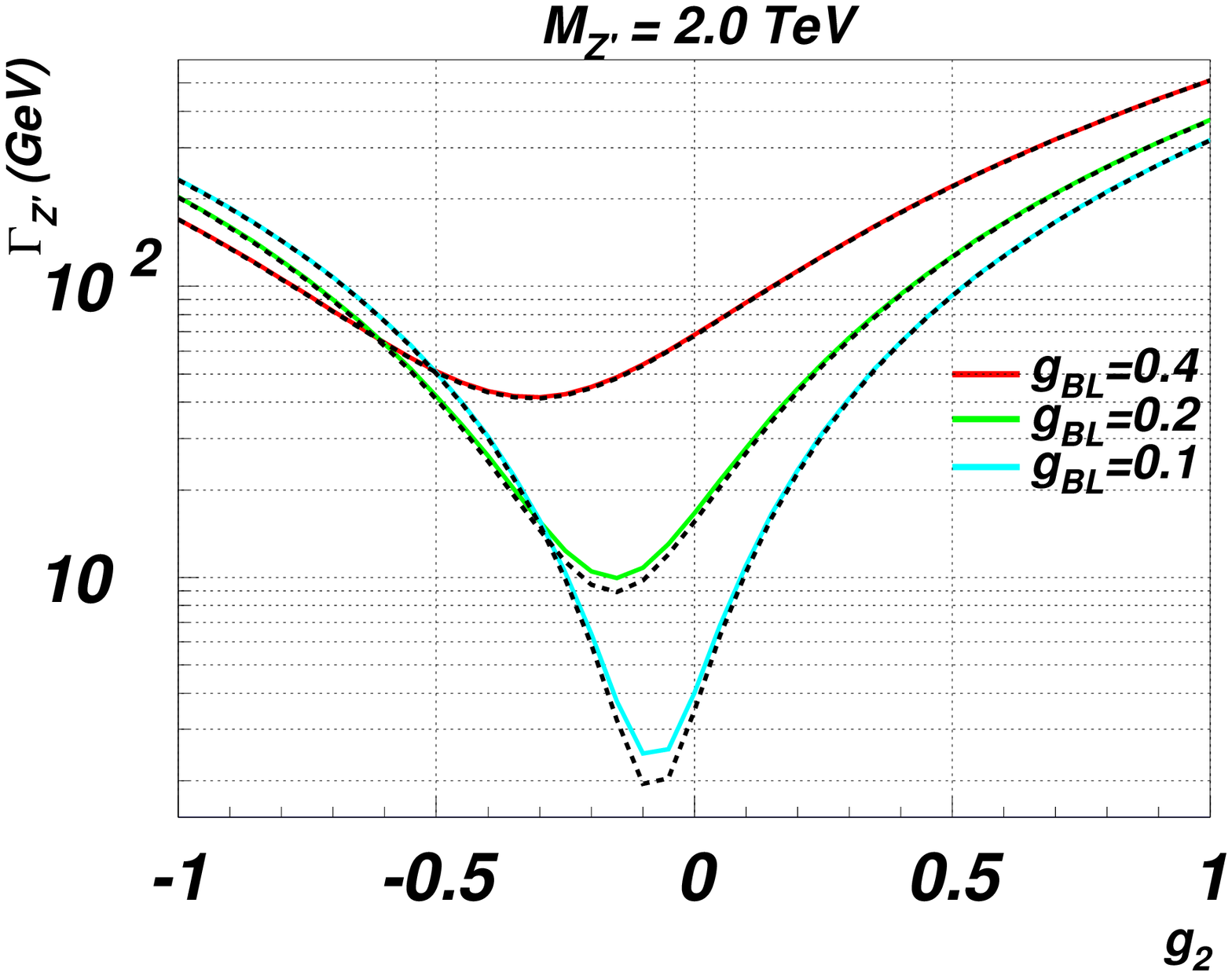}}
  \subfloat[]{
  \label{Pi_8_3}
  \includegraphics[angle=0,width=0.49\textwidth ]{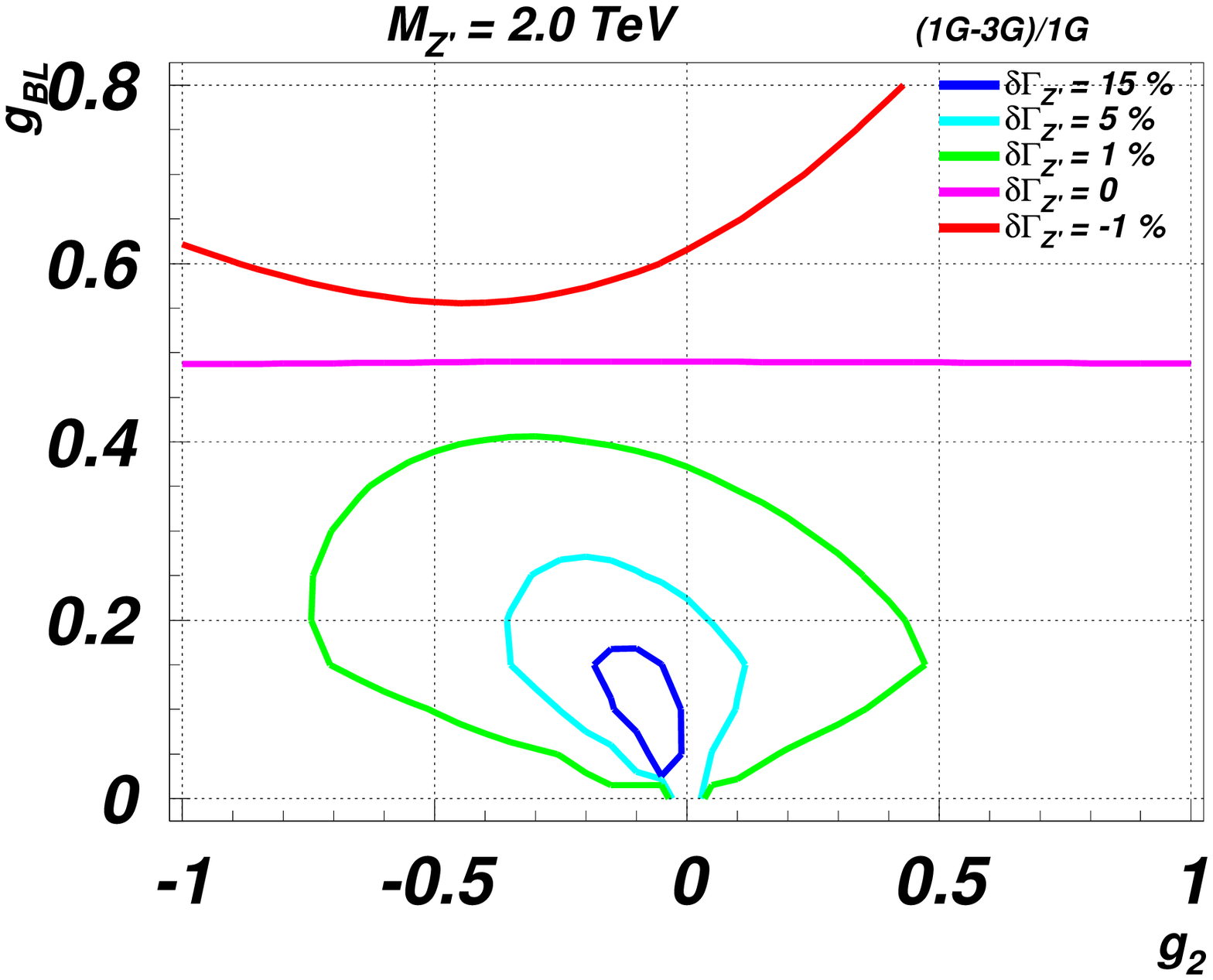}} 
\caption{(a) Total $Z'$ width for selected $g_{BL}$ values (solid lines refer to $N_\ell=1$, dashed lines to $N_\ell=3$) and (b) percentage variation between the $2$ models in the $g_{BL}-g_2$ plane, for $M_{Z'}=2.0$ TeV. For the heavy neutrino masses, see the text. \label{width}}
\end{figure}

We turn now to study the effect of having $3$ generations of leptinos, as required for leptogenesis, on the dark matter relic abundance due to the extra heavy neutrinos. The model with $3$ generations of leptinos has $3$ light and $6$ heavy Majorana neutrinos in total. The one with only $1$ generation of leptinos instead accounts for $3$ light neutrinos, $2$ of them being Dirac particles, and $2$ heavy Majorana neutrinos.
Although the total number of relativistic degrees of freedom, commonly addressed as $g^\ast$, is basically unchanged, the proliferation of neutrinos affects the $Z'$ width, and this could impinge on the evaluation of the relic abundance at the $Z'$ resonance. Figure~\ref{width} shows the total $Z'$ width in the $2$ different cases, $N_\ell=1,3$.


Despite the larger numbers of possible final states into which the $Z'$ boson can decay  for $N_\ell=3$ with respect to $N_\ell=1$, the latter case has a larger $Z'$ boson partial width into light neutrinos since 2 of them are Dirac particles. Moreover, finite mass and threshold effects for the heavy neutrinos diminish their branching ratios. Altogether, this induces an almost exact compensation between the larger number of heavy neutrinos for $N_\ell=3$ and the larger partial width due to the $2$ Dirac light neutrinos for $N_\ell=1$. 
The relative variation of the total $Z'$ width is below $1\%$ in most of the parameter space, getting above $5\%$ only in a limited region around small values of the gauge couplings.
 In these plots, heavy neutrino masses are $m_{h\nu}=637$ GeV for $g_{BL}=0.2$, and scale inversely with the latter. When their mass is above $M_{Z'}/2$ (or slightly before due to threshold effects, i.e., here for $g_{BL}<0.15$), the $Z'\to N N$ ($N = \nu_h,\, \nu{_h'}$) channels are all suppressed or simply forbidden and the above mentioned compensation is not taking place. Even though 
the model with $1$ generation of leptinos has here a bigger $Z'$ width due to the larger partial widths into the light Dirac neutrinos, the excess is never above $20\%$  of the total $Z'$ width in the case of $3$ generations, and only in a tiny corner of the parameter space, where both gauge couplings are small.

As intimated, the variation of the total $Z'$ width is very small, mostly below $1\%$, so that the impact on the DM relic abundance is also negligible. This is because, 
the relic abundance scales with the square root of the $Z'$ width. A $1\%$ variation in the total width determines a $0.5\%$ variation of the relic density.

We can conclude that the DM study for $N_\ell =1$ of section~\ref{sect:DM} is independent of the particular value of $N_\ell$ if the leptinos are not mass degenerate.

\section{Further results}
\label{sect:leptogenesis}
We comment here on further possible implications of the model. First, the possibility of  leptogenesis is described, necessarily requiring the presence of more than one generation of leptinos.
Next we comment on how our model has an effect on the effective number of light degrees of freedom $N_{eff}$, measured in cosmology.
Extra degrees of freedom consist of the RH neutrinos, forming Dirac fermions with the LH counterparts when less than $3$ generations of leptinos are present.

\subsection{Leptogenesis}
As discussed in the introduction, the inverse seesaw mechanism is a suitable mechanism for a large $CP$ asymmetry even at the TeV scale due to the naturally small mass splitting between the heavier neutrino eigenstates, a result that is built in into the model and does not require fine tuning. Hence, it is a favourable model for the so-called resonant leptogenesis. Despite a large loop enhancement
due to resonance, large phases are still required in the off-diagonal terms of $y_D$ to obtain an $\mathcal{O}(1)$ $CP$ asymmetry. In the $N_\ell=1$ model discussed so far, the first 2 generations of light neutrinos require $\mathcal{O}(10^{-12})$ Yukawas, which in turn also means that the $CP$--violating decays of the heavy Majorana neutrino pair are similarly suppressed. Large phases but small masses can be achieved extending the seesaw mechanism to the other generations.

We show here that leptogenesis, compatible with neutrino masses and mixing, is possible in this model.
The $CP$ asymmetry is generated by the decays of the heavy neutrinos:
\begin{equation}\label{cp-eff}
\varepsilon _i = \frac{1}{8\pi} \sum _{j\neq i} \frac{\mbox{Im} \left[ (y_D y^\dagger_D)^2_{ij}\right]}{\sum_\beta [y^D_{i\beta}]^2}f^\nu_{ij}\, ,
\end{equation}
requiring large phases in the off-diagonal elements of the Dirac Yukawa matrix $y_D$. In our model, these large phases, compatible with neutrino data, are possible for $N_l \ge 2$. The lepton number violating loop factor $f^\nu_{ij}$, when quasi-degenerate heavy neutrino pairs are considered, is
\begin{equation}\label{cp-resonance}
f^\nu_{ij} = \frac{M^2_j-M^2_i}{(M^2_j-M^2_i)^2+(M_j\Gamma_j-M_i\Gamma_i)^2}\, .
\end{equation}
In the inverse seesaw case, $M_j\sim M_i$ and $\Gamma_j = \Gamma_i\equiv \Gamma$ are naturally recovered, so that $f^\nu_{ij}$ can easily be $\mathcal{O}(1)$. Notice that no fine tuning is required.

Once an asymmetry in the lepton sector is produced, electroweak sphaleron processes take place and move the asymmetry to the baryon sector
$\displaystyle \eta _B = \frac{28}{79}\eta _{B-L}$.
Altogether, the final baryon asymmetry can be written as
\begin{equation}
\eta _B \sim 10^{-2}\, \varepsilon _i\, \kappa_i \,(z\to\infty)\, ,
\end{equation}
where the $10^{-2}$ pre-factor accounts for the sphaleron efficiency and for photon dilution after recombination. The $CP$ asymmetry $\varepsilon _i$ was defined in eq.~(\ref{cp-eff}), while $\kappa_i$ is the efficiency factor obtained after solving the relevant Boltzmann equations:
\begin{eqnarray}
\kappa _i(z) \sim \int^z_{z0} dz' \frac{d N^{eq}_{N_i}(z')}{dz'}
\frac{D(K_i, z')}{D(K_i,z') + 4S_{Z'}N^{eq}_{N_i}(z')}  \times \mbox{exp}\left[ -\int^z_{z'}dz'' W_{ID}(K_i,z'') \delta_i^2\right]\, ,
\end{eqnarray}
with $\displaystyle \delta_i=\frac{\left| M_i - M_j\right|}{\Gamma}\ll 1$ suppresses considerably the inverse decay (ID) wash-out.

All the quantities are defined in Ref.~\cite{Blanchet:2010kw}. Particularly important is the $S_{Z'}$ term, the $Z'$ scattering processes, which induces a wash-out of the final asymmetry. It has been verified that these $Z'$--induced wash-out processes do not have a severe impact on the final baryon asymmetry. This is due to the nature of the inverse seesaw mechanism, that allows for large Yukawa couplings, overcoming the $Z'$ processes. A similar conclusion was reached in Ref.~\cite{Iso:2010mv}, in which the authors explicitly showed that $Z'$ processes do not spoil the leptogenesis if the Dirac Yukawa couplings are sufficiently large, as in the inverse seesaw case under examination.

In conclusion, following the similar case studied in Ref.~\cite{Blanchet:2010kw}, the Dirac Yukawa matrix $y_D$ contains all information required to study the leptogenesis, entering both in $\varepsilon_i$ and in $\Gamma$. It is sufficient that some off-diagonal elements in $y_D$ are large and complex for leptogenesis to be possible. For definiteness, we have verified that the choice of the matrix in Ref.~\cite{Blanchet:2010kw}, that for $N_l=3$ is a possibility in our setup, does yield the correct baryon asymmetry in our model. We stress again that a similar choice for $y_D$ compatible with a successful leptogenesis is possible also for $N_l=2$, even though only one off-diagonal element will be large in this case. This is in fact identical to $N_l=3$ in the 1-flavour approximation.

Hence, we have proven that a successful leptogenesis, compatible with the observed pattern of neutrino masses and mixing angles is possible in our model when at least two  generations of leptinos are present. However, the complete analysis of the leptogenesis in our model, as for instance the detailed comparison of $N_l=2$ and $N_l=3$ cases, or the impact of flavour effects outside the simple 1-flavour analysis, is beyond the scope of this paper and is left for future work.

\subsection{Impact on $N^{\nu}_{eff}$}

As a last application of our model, we describe  the possible implications on $N^{\nu}_{eff}$. This observable counts the relativistic energy content in the universe at the time of the last scattering surface in terms of an effective number of neutrino species.
Indications from cosmology result in the observed value of $N^{CMB}_{eff} = 4.56 \pm 0.75$ when combining WMAP, the Atacama Cosmology Telescope, baryonic acoustic oscillations data and the measurement of the Hubble parameter $H_0$~\cite{Dunkley:2010ge}.
 Notice that the SM LH neutrinos only would yield $N^{\nu}_{eff} \sim 3$. This mismatch is typically interpreted as an indication of the existence of  some extra relativistic degrees of freedom that effectively contribute as one unit of $N^{\nu}_{eff}$. In our setup, the only extra degrees of freedom that can be relativistic at the last scattering surface are the RH neutrinos, when they have only Dirac mass terms, i.e., for less than $3$ generations of leptinos, as described in section~\ref{sect:ext_n_fam}. On the other hand, as we have seen previously, a successful leptogenesis requires more than one generation of leptinos. If one would like to explain both leptogenesis and $\Delta N^{\nu}_{eff}$ in our model, less than three generations of leptinos should be considered, given that $N_\ell =3$ leads to an effective number $N^{\nu}_{eff}\sim 3$ as in the standard model.

An estimate of the impact on $N^{\nu}_{eff}$ is done as follows. First, the decoupling temperature and the contribution to $N^{\nu}_{eff}$ depend on the parameters $m_{Z'}$, $g_{BL}$, $g_2$, given that only the $Z'$ boson can keep the RH neutrinos in thermal equilibrium with electrons, as the RH neutrinos have vanishing hypercharge. Naively, the bigger the $Z'$ cross section, the lower the decoupling temperature, which in turns also means that less degrees of freedom are relativistic at decoupling. Overall, this increases the RH neutrino contribution to $N^{\nu}_{eff}$. Close to LEP exclusion limits, we get a minimum  decoupling temperature of $~380$ MeV. At this temperature the number of relativistic degrees of freedom is $20$ if the QCD phase transition takes place at $450$ MeV~\cite{pdg2012}.
This yields the highest value for $\Delta N^{\nu}_{eff}$ in our model: $0.18\, (3-N_\ell)$. For lower QCD phase transition temperatures, the numbers of relativistic degrees of freedom rapidly increases, suppressing the impact on $N^{\nu}_{eff}$ to the percent level or below.

Thus we see that although the model can contribute to $N^\nu_{eff}$, its impact is only marginal, especially if one would want to implement a successful leptogenesis. In fact, $N_\ell \geq 2$ is required by the latter, while $N_\ell \to 0$ maximises $\Delta N^\nu_{eff}$.

\section{Conclusions}
\label{sect:conclusions}
We constructed a simple model in the class of the minimal $Z'$ models, where the extra
$U(1)$ gauge field is coupled only to hypercharge and \bml. We added right-handed neutrinos for each generation
of fermions, as required by the absence of chiral anomalies in the theory.
Beyond this we added extra pairs of leptons with fractional lepton number, which we therefore called leptinos.
One of the leptino in the pairs is chosen to be even and the other one to be odd under an additional $Z_2$ charge.
We were able to construct an inverse seesaw mechanism
for neutrino masses. The mechanism is natural in the sense that all possible terms consistent with the
symmetries of the theory are present in the Lagrangian. 
 Nonetheless some of the entries in the neutrino mass matrix are zero, because of the
choice of the representations. The reason is the presence of a fractional
lepton charge, which is the new feature of the model.

{
The choice of charges in combination with the extended gauge sector leads automatically to the
inverse seesaw mass matrix. Renormalizability does not allow for other terms. This is a major improvement
over the existing literature, where one has to evoke radiative terms or non-renormalizable interactions.
The stabilisation of the zeros in the inverse seesaw mass matrix also requires a $Z_2$ symmetry to be present.
The lightest odd particle under this unbroken $Z_2$ symmetry is then a dark matter candidate. This particle is needed 
to make sure that the new gauged $U(1)$ current is anomaly free.
Finally, the naturally degenerate heavy neutrinos in the inverse seesaw allows for a successful resonant leptogenesis at the TeV scale to explain the observed matter-antimatter asymmetry.
}


The odd leptino is the candidate for dark matter, as it is weakly 
interacting, massive and stable. We have shown that the correct dark matter density can be generated
if the leptino is annihilated through a resonance by either the $Z'$-boson or the Higgs-boson related to the breaking of the \bml\, symmetry.
We studied limits on the parameters of the theory,
coming from LEP, hadron colliders and the dark matter abundance. 
We found that the limits are such, that the $Z'$-boson and extra Higgs-boson lie within the range 
of the LHC, though the full design luminosity might be needed. The cross sections
are too small for present direct search experiments for dark matter. Resonant leptogenesis is possible
in the presence of more than one pair of leptinos, but contains no particularly new features
compared to other models of leptogenesis.

In conclusion the model provides a very simple extension of the standard model, containing a number of desirable features like dark matter, leptogenesis and (inverse) seesaw { for a viable neutrino sector.}
At the same time, being a singlet extension, it does not lead to phenomenological problems, such as flavour changing neutral currents.



 The model we constructed  appears to be able to give
a realistic description of cosmological data.
However, it is not possible to describe at the same time leptogenesis and $N^\nu_{eff}$.
 { The model can be tested at the LHC, because the  new particles must have masses in the LHC range for the model to explain the cosmological observations. Furthermore, new signals arise in the heavy neutrino sector, see~\cite{delAguila:2008cj,Basso:2008iv,Hirsch:2009ra}. The characterisation of the signals in the inverse seesaw model as compared to the type-I case is subject of further investigations.}

\section*{Acknowledgements}
\label{Sec:acknowledgements}
This work is supported by the 
Deutsche Forschungsgemeinschaft through the Research Training Group grant
GRK\,1102 \textit{Physics at Hadron Accelerators}.

\appendix
\label{appendix}
\section{Gauge couplings RGEs}

We present here the 
renormalization group equations for the Abelian gauge couplings. The one-loop RGEs read~\cite{delAguila:1988jz,delAguila:1995rb,Ferroglia:2006mj}
\begin{eqnarray}\label{RGE_g1}
\frac{d}{dt}g_1 &=& \frac{1}{16\pi ^2}\left[A^{YY}g_1^3 \right]\, , \\ \label{RGE_gBL}
\frac{d}{dt}g_{BL} &=& \frac{1}{16\pi ^2}\left[A^{XX}g_{BL}^3+2A^{XY}g_{BL}^2g_2+A^{YY}g_{BL}g_2^2 \right] \, , \\ \label{RGE_g2}
\frac{d}{dt}g_2 &=& \frac{1}{16\pi ^2}\left[A^{YY}g_2\,(g_2^2+2g_1^2)+2A^{XY}g_{BL}(g_2^2+g_1^2)+A^{XX}g_{BL}^2g_2 \right]\, ,
\end{eqnarray}
For the model we are discussing ($Y$ is the SM weak hypercharge, $X=$\bml\, is the \bml\, number), the coefficients are:
\begin{equation}
A^{YY}=\frac{41}{6}\, ,\qquad A^{XX}=\frac{32+(Y^{\bml}_{\chi})^2}{3}+\frac{4}{27}N_\ell\, ,\qquad A^{YX}=\frac{16}{3}\, ,
\end{equation}
if $N_\ell$ generations of leptinos (section C) are included. Notice the small difference with
respect to Ref.~\cite{Ferroglia:2006mj} in the $A^{YY}$ coefficient, due to the SM Higgs boson in the counting. 

{
The gauge boson mixing is controlled by eq.~(\ref{RGE_g2}), that is the evolution of the mixing gauge coupling $g_2$. Even if at the EW scale such mixing is set to vanish, the one-loop running will induce it because the equation for $g_2$ is not proportional to $g_2$ itself. Equivalently, the one-loop-induced gauge mixing can be included at the EW scale by a suitable choice of the $g_2$ mixing coupling. 

The study of the RGEs for the gauge couplings is important to set upper limits on $g_2$ and $g_{BL}$ couplings at the EW scale to avoid Landau poles somewhere up to the Plank scale, see Ref~\cite{Basso:2010jm}.} 

The equations can be solved algebraically~\cite{Ferroglia:2006mj}:
\begin{eqnarray}\label{soleq_1}
\frac{1}{g_1^2}+2A^{YY}t &=& \mbox{constant}\, ,\\\label{soleq_2}
\frac{2A^{YY}g_2+2A^{YX}g_{BL}}{g_1^2g_{BL}} &=& \mbox{constant}\, ,\\\label{soleq_3}
\frac{A^{YY}(g_1^2+g_2^2)-A^{XX}g^2_{BL}}{g_1^2g_{BL}} &=& \mbox{constant}\, ,
\end{eqnarray}
Particularly interesting is eq.~(\ref{soleq_2}), which leads to an infrared (IR) fixed point for the model
\begin{equation}\label{IRfix}
41g_2+32g_{BL}=0\, .
\end{equation}
This fixed point is independent of the additional matter we consider in the model. The reason for this is that the extra fields we introduce (RH neutrinos, leptinos and the singlet scalar) are all singlets under the SM gauge group, hence not entering in the diagonal and mixed hypercharge coefficients $A^{YY}$ and $A^{XY}$,  that are the only terms appearing in eq.~(\ref{soleq_2}). In other words, this IR fixed point is a model independent property of the minimal $Z'$ model.

\bibliography{biblio}

\end{document}